\documentclass[apj]{emulateapj}

\slugcomment{Accepted for publication in ApJ 2015 April 21; published 2015 June 10.}

\usepackage{amssymb,amsmath}
\usepackage{mathrsfs}
\usepackage{natbib}
\usepackage{mathtools}
\usepackage{graphicx}
\usepackage{multirow}
\usepackage{subfigure}
\usepackage{eucal}
\usepackage{url}
\usepackage{color}

\newcommand{\mb}{\ensuremath{\mathrm{M_\star^B }}}
\newcommand{\md}{\ensuremath{\mathrm{M_\star^D }}}
\newcommand{\mt}{\ensuremath{\mathrm{M_\star }}}
\newcommand{\sfr}{\ensuremath{\mathrm{\dot{M}_\star }}}
\newcommand{\massunits}{\ensuremath{\mathrm{M_\odot }}}
\newcommand{\sfrunits}{\ensuremath{\mathrm{M_\odot}} \ensuremath{\mathrm{yr^{-1}}}}

\newcommand{\mbi}{\ensuremath{\mathrm{M_{\star,i}^B}}}
\newcommand{\mdi}{\ensuremath{\mathrm{M_{\star,i}^D}}}
\newcommand{\mti}{\ensuremath{\mathrm{M_{\star,i}}}}

\newcommand{\D}{\ensuremath{\mathcal{D}}}
\newcommand{\Ob}{\ensuremath{\mathcal{O}}}
\newcommand{\N}{\ensuremath{\mathcal{N}}}

\newcommand{\cobe}{\emph{COBE}}

\begin{document}

\title{Improved Estimates of the Milky Way's Stellar Mass and Star Formation Rate from Hierarchical Bayesian Meta-Analysis}
\author{Timothy C. Licquia$^{1,2}$, Jeffrey A. Newman$^{1,2}$}
\affil{$^1$
Department of Physics and Astronomy, University of Pittsburgh, 
3941 O'Hara Street, Pittsburgh, PA, 15260; tcl15@pitt.edu\\
$^2$
PITTsburgh Particle physics, Astrophysics, and Cosmology Center (PITT PACC)}


\begin{abstract}

We present improved estimates of several global properties of the Milky Way, including its current star formation rate (SFR), the stellar mass contained in its disk and bulge+bar components, as well as its total stellar mass.  We do so by combining previous measurements from the literature using a hierarchical Bayesian (HB) statistical method that allows us to account for the possibility that any value may be incorrect or have underestimated errors.  We show that this method is robust to a wide variety of assumptions about the nature of problems in individual measurements or error estimates.  Ultimately, our analysis yields a SFR for the Galaxy of $\sfr=1.65\pm0.19$ \sfrunits, assuming a Kroupa initial mass function (IMF).  By combining HB methods with Monte Carlo simulations that incorporate the latest estimates of the Galactocentric radius of the Sun, $R_0$, the exponential scale length of the disk, $L_d$, and the local surface density of stellar mass, $\Sigma_\star(R_0)$, we show that the mass of the Galactic bulge+bar is $\mb=0.91\pm0.07\times10^{10}$ \massunits, the disk mass is $\md=5.17\pm1.11\times10^{10}$ \massunits, and their combination yields a total stellar mass of $\mt=6.08\pm1.14\times10^{10}$ \massunits\ (assuming a Kroupa IMF and an exponential disk profile).  This analysis is based upon a new compilation of literature bulge mass estimates, normalized to common assumptions about the stellar initial mass function and Galactic disk properties, presented herein. We additionally find a bulge-to-total mass ratio for the Milky Way of $B/T=0.150^{+0.028}_{-0.019}$ and a specific star formation rate of $\sfr/\mt=2.71\pm0.59\times10^{-11}$ yr$^{-1}$.

\end{abstract}


\section{Introduction} \label{sec:intro}
Determining the global properties of the Milky Way inherently poses unique challenges.  Unlike any other galaxy in the universe, we lack the ability to study the Milky Way from an outside perspective.  This disadvantage is greatly amplified by our location within the disk, forcing us to peer through the dusty interstellar medium (ISM) when looking toward other stars.  The Galactic ISM inhibits our view of more distant regions of the disk, particularly in the optical and near-UV wavelengths \citep[cf.][]{Herschel,SFD}.  For these reasons, there are a limited number of studies in the literature that aim to produce a global picture of our Galaxy.

Here, we aim to improve our understanding both of the total star formation rate (SFR) and the total stellar mass of the Milky Way.  We do this, \textit{not} by analyzing any new observational data, but by statistically combining the prior measurements of these properties in the literature using the power of a hierarchical Bayesian (HB) method \citep{Press}.  Such methods are not new to astronomy, though still rare in the literature; they can be a robust and versatile tool for both data and model analysis, and subsequently their prevalence in the literature has grown greatly in the past few decades \citep{Loredo}.  For instance, \citet{Newman99} applied this technique (in a maximum likelihood framework) to combine the observed properties of $43\pm7$ real and similar number of imposter Cepheid variables found in the Centaurus cluster, handling the possibility of false positive detections, in order to determine a period-luminosity-relation based distance modulus to NGC 4603. \citet{LangHogg} were able to produce tight constraints on the orbit of Comet 17P/Holmes by applying this method to the diverse set of image query results for ``Comet Holmes'' obtained from the \textit{Yahoo!} internet search engine.  \citet{March2014} present an HB model to improve constraints on cosmological parameters by combining information from supernovae (SN) Ia lightcurves.  \citet{ShettyA,ShettyB} use the HB method to reveal the Kennicutt-Schmidt relation, i.e., the relationship between star formation rate, \sfr, and molecular gas surface density, $\Sigma_{\mathrm{mol}}$, to be non-universal and in many cases sub-linear, indicating that $\Sigma_{\mathrm{mol}}$ alone is insufficient to predict a galaxy's SFR.  Most recently, \citet{Mandel} applied this method in order to disentangle the effect of systematic reddening due to host galaxy dust from the expansion-velocity-dependent variation in intrinsic SN Ia colors.

Adopting a Bayesian perspective, our major goal in this paper is to answer the question: Given the previous measurements of a given parameter of the Milky Way, what conclusions can we draw about its true value?  We first apply this analysis method to estimate the Milky Way's SFR.  We next consider the bulge and overall mass of the Milky Way.  This introduces additional complications due to variations in the Galactocentric radius of the Sun, $R_0$, assumed in different measurements.  To deal with this, we combine our HB analysis with Monte Carlo (MC) simulations which incorporate the current uncertainties in $R_0$.  The MC method allows us to simultaneously produce new estimates for the stellar mass contained in the bulge+bar, \mb, the stellar disk mass, \md, and the total stellar mass, \mt, of the Milky Way, assuming the single-exponential disk model from \citet{Bovy2013} and incorporating uncertainties in $R_0$, the exponential scale length of the disk, $L_d$, and the local surface density of stellar material, $\Sigma_\star(R_0)$.

We structure this paper as follows.  In \S\ref{sec:bayes} we describe the hierarchical Bayesian formalism we use in order to construct aggregate results incorporating the information from a variety of independent measurements.  In \S\ref{sec:sfr} we apply this technique to the prior estimates of the Milky Way's star formation rate.  The results of the SFR analysis are discussed in \S\ref{sec:sfr_results}.  Next, we take on a more complex example, ultimately constraining the total stellar mass in the Milky Way.  \S\ref{sec:mstar_diskmodel} details the stellar disk model we assume for this study.  In \S\ref{sec:monte_carlo}, we explain how we use Monte Carlo simulations to produce a new estimate of the disk mass, to supplement the HB analysis used to determine the bulge+bar mass, and we combine these two results to yield the total stellar mass in the Galaxy.  The results for these three measurements are discussed in \S\ref{sec:mstar_results}.  We summarize and discuss the main results of this study in \S\ref{sec:summary}.  Lastly, in the Appendix we explore how plausible it is that Galactic disk deviates from a pure exponential profile, and investigate how this would affect our results.

\section{Hierarchical Bayesian Analysis} \label{sec:bayes}
In this section, we describe the analysis methods we use to combine a set of measurements for some observable (e.g., \sfr\ or \mb), along with their attendant uncertainties, into one aggregate result using a hierarchical Bayesian (HB) formalism.  Ultimately, this process provides a new probability distribution function (PDF), referred to in Bayesian terms as the \textit{posterior}, for the observable of interest by combining the PDFs yielded from multiple individual estimates, enabling us to incorporate the information gained from a variety of independent observations and analyses.  This method allows us to account for the possibility that any of the measurements may be incorrect, or affected by systematic errors that have been overlooked.  This is a key advantage over calculating simpler statistics, such as the standard inverse-variance weighted mean, which are more easily skewed by outliers and are also contingent on the assumptions that the individual PDFs are Gaussian in form and statistically compatible.

\subsection{Defining the Problem} \label{sec:understanding_data}
We begin by collecting a set of $N$ independent measurements of observable \Ob\ from the literature, and denote this dataset as \D.  In a Bayesian framework, each study can be considered to provide a PDF for the true value of \Ob, $\mu_0$, given the data obtained; if the probability distribution is Gaussian in form, it can be described by its mean value $\mu_i$ and standard deviation $\sigma_i$.  Under the assumption of normality, we know there should be a $\sim$68\% chance of $\mu_0$ being in the range $\mu_i\pm\sigma_i$ and $\sim$95\% chance of it being within $\mu_i\pm2\sigma_i$, assuming all errors (statistical and systematic) are accounted for in $\sigma_i$.  Of course, this is not always a safe assumption; often we may find that two separate measurements of \Ob\ are in tension with each other, producing results with $1\sigma$ or even $2\sigma$ confidence regions that do not overlap.  If this tension is sufficiently great, we can conclude that \textit{at least} one of the estimates of $\mu_0$ must be affected by a bias or systematic error that has not been incorporated in $\sigma_i$; this is not sufficient to determine which of the two estimates is problematic.

\subsection{Relieving the Tension} \label{sec:relieving_tension}
Practically speaking, the measurements included in our dataset, given their nominal errors, must always be in tension with each other at some level of significance.  In order to resolve this tension we can hypothesize that some of these studies have overestimated their ability to measure \Ob.  Suppose that $f_{\text{good}}$ denotes the fraction of ``good'' measurements; i.e., the probability that any single measurement within \D\ is accurately described by $\mu_i$ and $\sigma_i$.  Thus, $1-f_{\text{good}}$ is the global fraction of measurements that are ``bad'' --- i.e., inaccurate --- generally due to underestimated error bars (e.g., ones that omit the possibility of significant systematic errors).  With no \textit{a priori} knowledge of which estimates are not ``good'', we must find a way to remedy their effect when obtaining combined constraints on $\mu_0$.  In this study we explore a number of ways to do this.  For instance, perhaps the ``bad'' estimates simply require their error bars to be scaled by a constant factor; we will denote the resulting degraded uncertainty estimate by $\sigma_{n,i}=n\sigma_i$.  We refer to this as the ``free-$n$'' model below.

More likely, problematic measurements may have overlooked systematic uncertainties in their methods which should be added in quadrature to the nominal error estimates.  Let $\mu_i^{\textrm{MED}}$ denote the median of all the $\mu_i$ values.  We investigate what happens when adding a fractional amount $Q$ of $\mu_i^{\textrm{MED}}$ in quadrature to the nominal error bars, such that the error for a ``bad'' estimate is given by $\sigma^2_{Q,i}=\sigma_i^2 + (Q\mu_i^{\textrm{MED}})^2$.  We use $Q\mu_i^{\textrm{MED}}$ here, instead of $Q\mu_i$, to avoid a slight bias toward assigning lower errors to lower valued estimates of $\mu_0$.  We refer to this as the ``free-$Q$'' model below.  Instead, we could consider a case where there is a floor value of fractional error which the the nominal error bars should not dip below.  In such a case we would use $\sigma_{F,i}=F\mu_i^{\textrm{MED}}$ if $\sigma_i<F\mu_i^{\textrm{MED}}$, or otherwise $\sigma_{F,i}=\sigma_i$.  We refer to this as the ``free-$F$'' model below.

Furthermore, it is possible that some of the included estimates are entirely wrong, and thus should effectively be excluded from \D.  In this scenario we would handle this by replacing the nominal Gaussian PDF with $\mu_i$ and $\sigma_i$ by a uniform probability distribution over the full potential parameter range.  We label this as the ``$P_\text{bad}$-flat'' model below.  We also examine the results of assuming we have included no ``bad'' measurements in our analysis by forcing $f_{\text{good}}=1$ instead of allowing it as a free parameter (i.e., to assume that all of the $\mu_i$ and $\sigma_i$ are correct).  All of these ways of dealing with the inclusion of ``bad'' measurements in our dataset have different advantages and address the problem from a slightly different approach.  A hierarchical model allows us to quantify the affect of the ``bad'' measurements on the combined results by simultaneously fitting parameters that describe the data (e.g., $f_{\text{good}}$, $n$, $Q$, $F$), while also fitting for those that describe the physical model (e.g., $\mu_0$).  It is important to note that, in all cases, the constraints on the physical parameter of interest yielded by this technique will only be improved if the systematics affecting the individual measurements are different, as systematic errors that are common across a set of measurements do not improve by adding more data.  In the following section, we describe the HB formalism we use to analyze each scenario and the criteria we use to distinguish which of these models best fits the data.

\subsection{The Formalism} \label{sec:formalism}
We closely follow the prescription laid forth by \citet{Press} and refer the reader to this work for a more in-depth derivation.  Bayes' theorem provides the framework in which we can calculate the probability of a particular model given data \D: 
\begin{equation} \label{eq:bayes}
P({\Theta \mid \D}) = \frac{P({\D \mid \Theta})P(\Theta)}{P(\D)} ,
\end{equation}
where $\Theta$ is a vector containing the free parameters of the model.  Here, the \textit{posterior} probability $P({\Theta \mid \D})$ is equal to the product of the \textit{likelihood} $P({\D \mid \Theta})$ and the \textit{prior} $P(\Theta)$, divided by the \textit{evidence} $P(\D)$.  The likelihood is the probability of obtaining the actual measurements $\D$, given that the parameters $\Theta$ specify the correct model of the data.  The prior reflects our previous knowledge of what the parameters of the true model are, before the data \D\ are considered.  In general, this must be subjectively chosen.  The evidence represents the overall probability of finding the data in hand, and when considered on its own it provides a useful means of comparing different models, as we will discuss in \S\ref{sec:better_model}.  This is obtained by integrating the likelihood weighted by the prior (i.e., the numerator of Equation \eqref{eq:bayes}) over all possible values of $\Theta$; hence it is also sometimes called the \textit{marginal likelihood}.  As written here, the posterior yields a properly normalized PDF representing the degree of belief of the model parameters $\Theta$ being in a given range.

\subsubsection{The Likelihood} \label{sec:likelihood}
We begin with the assumption that all measurements may be represented by the combination of two probability distributions, representing the possibility that it is ``good'' or ``bad''.  The PDF for $\mu_0$ given a measurement will be $P(\mu_0) = P({\mu_0 \mid \text{``good''}})P(\textrm{``good''})+P({\mu_0 \mid \textrm{``bad''}})(1-P(\textrm{``good''}))$; the probability that a given measurement is ``good'' is simply $f_{\text{good}}$.  We assume that each measurement in \D\ is statistically independent from all the others; in that case we may write the overall likelihood for \D\ as the product of the likelihoods for each measurement we include.  For each of the scenarios described above in \S\ref{sec:relieving_tension}, the appropriate likelihood is given by:

For the free-$n$ model (i.e., where ``bad'' measurements are assumed to be underestimating errors by a constant factor), the likelihood is \footnotesize
\begin{align}
P({\D \mid \mu_0, f_{\text{good}},n}) &= \displaystyle\prod\limits_{i=1}^N \Bigg(\frac{f_{\text{good}}}{\sqrt{2\pi\sigma_i^2}}\exp\bigg[\frac{-(\mu_i-\mu_0)^2}{2\sigma_i^2}\bigg] \nonumber \\
&\quad\quad+\frac{1-f_{\text{good}}}{\sqrt{2\pi(n\sigma_i)^2}}\exp\bigg[\frac{-(\mu_i-\mu_0)^2}{2(n\sigma_i)^2}\bigg]\Bigg).
\label{eq:likelihood_n}
\end{align}
\normalsize

For the free-$Q$ model (i.e., where ``bad''-measurement errors are assumed to require extra uncertainty added in quadrature), the likelihood is \footnotesize
\begin{align}
&P({\D \mid \mu_0, f_{\text{good}},Q}) =  \nonumber \\
& \quad\displaystyle\prod\limits_{i=1}^N \Bigg(\frac{f_{\text{good}}}{\sqrt{2\pi\sigma_i^2}}\exp\bigg[\frac{-(\mu_i-\mu_0)^2}{2\sigma_i^2}\bigg] \nonumber \\
& \quad\quad+\frac{1-f_{\text{good}}}{\sqrt{2\pi\Big((\sigma_i)^2+(Q\mu_i^{\textrm{MED}})^2\Big)}}\exp\bigg[\frac{-(\mu_i-\mu_0)^2}{2\big((\sigma_i)^2+(Q\mu_i^{\textrm{MED}})^2\big)}\bigg]\Bigg).
\label{eq:likelihood_Q}
\end{align}
\normalsize

For the free-$F$ model (i.e., where ``bad'' measurements are assumed to be underestimating errors only if they are below a minimum threshold), the likelihood is \footnotesize
\begin{align}
&P({\D \mid \mu_0, f_{\text{good}},F}) = \nonumber \\
& \quad\quad \displaystyle\prod\limits_{i=1}^N 
		\begin{cases}
			\frac{f_{\text{good}}}{\sqrt{2\pi\sigma_i^2}}\exp\left[\frac{-(\mu_i-\mu_0)^2}{2\sigma_i^2}\right] \\
			\quad+\frac{1-f_{\text{good}}}{\sqrt{2\pi(F\mu_i^{\textrm{MED}})^2}}\exp\left[\frac{-(\mu_i-\mu_0)^2}{2(F\mu_i^{\textrm{MED}})^2}\right],& \text{if } \sigma_i<F\mu_i^{\textrm{MED}}\\
			\frac{1}{\sqrt{2\pi\sigma_i^2}}\exp\left[\frac{-(\mu_i-\mu_0)^2}{2\sigma_i^2}\right],& \text{otherwise.} \label{eq:likelihood_F}\\
		\end{cases}
\end{align} \normalsize
In this model, $f_{\text{good}}$ is the fraction of ``good'' measurements assuming the given value of $F$; i.e., it is the fraction of accurate measurements amongst those with $\sigma_i<F\mu_i^{\textrm{MED}}$.  This is subtly different from the former two models, where $f_{\text{good}}$ characterizes the entire dataset at any given value of $n$ or $Q$.

For each of the above 3 models, we also explore the results of assuming that \textit{all} measurements included in \D\ have misestimated errors to some extent by \textit{not} treating $f_{\text{good}}$ as a free parameter, but rather setting it to zero.

For the $P_\text{bad}$-flat model (i.e., where ``bad'' measurements are assumed to be entirely wrong and discardable, and so we replace their PDFs with a flat distribution), the likelihood is \footnotesize
\begin{align}
P({\D \mid \mu_0, f_{\text{good}}}) & = \displaystyle\prod\limits_{i=1}^N \bigg(\frac{f_{\text{good}}}{\sqrt{2\pi\sigma_i^2}}\exp\left[\frac{-(\mu_i-\mu_0)^2}{2\sigma_i^2}\right] \nonumber \\
 &\quad\quad\quad\quad\quad\quad\quad+(1-f_{\text{good}})\times\text{const.}\bigg).
\label{eq:likelihood_pbad_flat}
\end{align} \normalsize
The constant here is chosen such that the integral of $P({\mu_0 \mid \textrm{``bad''}})$ is equal to 1.

Lastly, there is the possibility that all of the measurements included in our analysis are ``good'' and have accounted for all uncertainties in their analyses.  We can determine $\mu_0$ for this scenario by setting $f_{\text{good}}$ to unity; in this case, the results of the HB analysis method become equivalent to the inverse-variance weighted mean of the individual measurements.  Effectively, this model serves as the null hypothesis of our study, and we denote this as the all-``good'' model when comparing our results.

In total, this yields 8 different ways of modeling ``bad'' measurements that could influence our aggregate estimate of $\mu_0$.  In \S\ref{sec:better_model} we will describe which of these options is best to follow.

\subsubsection{The Prior} \label{sec:priors}
If we assume that our prior knowledge of the parameters of our model are unrelated to each other, the overall prior for $\Theta$ is separable into the product of the priors for each parameter.  In example, in the context of this paper, we would not expect that the probability of an astronomer quoting accurate error bars on his/her result should be different if the Milky Way is truly producing 3 solar masses worth of new stars each year as opposed to 2.  This means we can write the joint prior $P(\mu_0,f_{\text{good}})$ as $P(\mu_0)P({f_{\text{good}} \mid \mu_0})=P(\mu_0)P(f_{\text{good}})$.  Likewise, in the absence of data, we would not expect a parameter that describes how inaccurate a study's error bars are to be dependent on the probability of any one measurement being ``bad'' or the mass of new stars being formed in the Galaxy per year.

For all 4 free parameters we use to characterize ``good'' and ``bad'' estimates we choose flat priors, meaning that we believe there is a 100\% chance of the true value being within a given range, and any single value within that range has equal probability to any other.  We note that these assumptions are somewhat arbitrary; for instance, a flat prior in a given quantity corresponds to a non-flat prior for the log of that quantity.  Flat priors in log space are generally preferred for quantities with no relevant scales/order of magnitude; however, that does not apply here.  Effectively, our choice causes all posterior distributions we calculate to be determined only by the likelihood of the observed data.  For the free-$n$ model we assume 100\% probability that $n$ is in the range [1,4]; i.e., that the errors for ``bad'' measurements will be underestimated by a factor no less than 1 and no greater than 4.  In turn, for the free-$Q$ model we assume 100\% probability that $Q$ is in the range [0,1]; i.e., the extra error needed to be added in quadrature to correct the nominal error bars of a ``bad'' measurement is no less than 0 and no more than 100\% of $\mu_i^{\textrm{MED}}$.  Lastly, for the free-$F$ model we assume 100\% probability that $F$ is in the range [$\sigma_i^{\textrm{MIN}}/\mu_i^{\textrm{MED}}$,1], where $\sigma_i^{\textrm{MIN}}$ is the minimum error estimate of all measurements included in \D; i.e., the minimum error estimate for any ``bad'' measurement should be no less than $\sigma_i^{\textrm{MIN}}$ and no larger than $\mu_i^{\textrm{MED}}$.  The priors can then be expressed as piece-wise functions,

\begin{equation}
P(f_{\text{good}})=
	\begin{cases}
	1,	& \text{if } 0 \leq f_{\text{good}} \leq 1\\
	0,	& \text{otherwise;}\\
	\end{cases} 
\end{equation}

\begin{equation}
P(n)=
	\begin{cases}
	\frac{1}{3},	& \text{if } 1 \leq n \leq 4\\
	0,	& \text{otherwise;}\\
	\end{cases} 
\end{equation}

\begin{equation}
P(Q)=
	\begin{cases}
	1,	& \text{if } 0 \leq Q \leq 1\\
	0,	& \text{otherwise;}\\
	\end{cases} 
\end{equation}

\begin{equation}
P(F)=
	\begin{cases}
	\frac{1}{1-\sigma_i^{\textrm{MIN}}/\mu_i^{\textrm{MED}}},	& \text{if } \sigma_i^{MIN}/\mu_i^{\textrm{MED}} \leq F \leq 1\\
	0,	& \text{otherwise.}\\
	\end{cases} 
\end{equation}

\subsubsection{The Marginalized Posterior} \label{sec:marginalizaiton}
In this subsection, we detail how the posterior PDF for a single parameter of our model is produced from the joint posterior from the HB analysis, using the free-$n$ model as an example.  Returning to Equation \eqref{eq:bayes}, the posterior PDF for the parameters of our model will be \footnotesize
\begin{align}
&P({\mu_0, f_{\text{good}},n \mid \D}) \propto \nonumber \\
&\,P(\mu_0)P(f_{\text{good}})P(n) \displaystyle\prod\limits_{i=1}^N \bigg(\frac{f_{\text{good}}}{\sqrt{2\pi\sigma_i^2}}\exp\Big[\frac{-(\mu_i-\mu_0)^2}{2\sigma_i^2}\Big]\nonumber\\
&\quad\quad\quad\quad\quad\quad\quad\quad\quad\quad+\frac{1-f_{\text{good}}}{\sqrt{2\pi(n\sigma_i)^2}}\exp\Big[\frac{-(\mu_i-\mu_0)^2}{2(n\sigma_i)^2}\Big]\bigg).
\end{align} \normalsize
It is often more informative to consider the posterior for an individual parameter; we can simply obtain this by \textit{marginalizing}: i.e., integrating the PDF over all other parameters of the model.  For instance, in our example, if we are interested in the marginalized posterior for $\mu_0$, then we calculate
\begin{equation}
P(\mu_0 \mid \D) = \int \int P({\mu_0,f_{\text{good}},n \mid \D})\,\mathrm{d}f_{\text{good}}\,\mathrm{d}n.
\label{eq:post_mu_0}
\end{equation}
Note that this result averages over all possible values of $f_{\text{good}}$ and $n$.  Lastly, we normalize the posterior to be a true PDF by dividing by the evidence, which is obtained by integrating Equation \eqref{eq:post_mu_0} over all $\mu_0$.

\subsubsection{Choosing Amongst Models} \label{sec:better_model}
In this study, we consider a variety of ways to model the inclusion of ``bad'' measurements in our dataset.  Generally speaking, a model with more free parameters will be able to produce a better fit to the data; however, the constraints on the model will be weaker, as there are more degeneracies introduced between parameters.  In order to compare the utility of the models to each other, we calculate the \textit{evidence} for each model.  If $M_i$ labels a particular model which is specified by parameters $\Theta$, then the evidence for model $M_i$ is 
\begin{align}
P({\D \mid M_i}) & = \int P({\D, \Theta \mid M_i})\,\mathrm{d}\Theta \nonumber \\
& = \int P({\D \mid \Theta}) P({\Theta \mid M_i})\,\mathrm{d}\Theta;
\end{align}
this is simply the likelihood integrated over all possible parameter values of the model, weighted by their priors.  This provides a natural, Bayesian method of incorporating the principle of Ockham's razor into our model comparison.  Essentially, the evidence will be maximized by a model's ability to better fit the data; however, this will be compromised if excessive parameter space is required to achieve such a fit.  As the all-``good'' model represents the null hypothesis for our HB analysis, for each model we report the \textit{Bayes Factor} defined as
\begin{equation}
K = \frac{P({\D \mid M_i})}{P({\D \mid M_{\text{all-``good''}})}}.
\end{equation}
Effectively, $K$ represents the ratio of the posterior odds to the prior odds of $M_i$ the being the correct model over $M_{\text{all-``good''}}$ \citep[][and references therein]{Kass}. As defined this way, we choose the best model of the data to be the one with largest $K$.  In our tables we list for each model the difference in $\log_{10} K$ compared to our fiducial model; a difference of 2 is commonly used to indicate statistically significant differences in model utility.

As alternatives to the Bayes Factor, we also report the Akaike information criterion \citep[][hereafter AIC]{Akaike}, defined as
\begin{equation}
\text{AIC} = -2\ln\mathscr{L}+2k,
\end{equation}
and the Bayesian information criterion \citep[][hereafter BIC]{Schwarz1978}, defined as
\begin{equation}
\text{BIC} = -2\ln\mathscr{L}+k\ln N,
\end{equation}
where $N$ is the number independent measurements included in our analysis, $k$ is the number of free parameters of the model, and $\mathscr{L}$ is the maximum likelihood value.  These provide a secondary and less sophisticated way of weighing the goodness of fit against the number of free parameters included in each model, and whose comparison can serve as a rough (and less computationally intensive) approximation to comparing $\log_{10} K$ values.  However, in contrast to the $K$ values, the best model of the data would be the one that \textit{minimizes} the information criteria.  Thus, for each model we report $\Delta$AIC (or $\Delta$BIC) measured from the \textit{lowest} AIC (or BIC) value amongst all models; similar to the $\Delta\log_{10} K$ values, a change of $\sim$2 is indicative of statistically significant differences in model utility \citep{burnham2002}.  Note that, as AIC and BIC values are calculated on a natural logarithm scale, this constraint is a bit less conservative than the one used for $\Delta\log_{10} K$ values (i.e., bigger differences in $\Delta\ln K$ are required to indicate a significant difference).

The remainder of this paper focuses on two application of the HB method.  First, we use this technique to constrain the star formation rate of the Milky Way.  In our second and more complex example, we produce an hierarchical estimate of the stellar mass contained in the combined bulge \& bar components of the Milky Way, where we must incorporate Monte Carlo simulations into our technique to reflect uncertainties in the Sun's Galactocentric radius.  Simultaneously, we produce PDFs for the stellar mass of the disk component of the Galaxy, as well as its total stellar mass.  As mentioned earlier, we are effectively assuming that if there are systematic errors in measurements of the Milky Way properties, they will generally not be in common amongst all the techniques, but rather, given the multitude of methods applied, many methods should have different systematics (of differing signs).  If that is not the case, errors will not be reduced when combining information from multiple results.

\section{The Milky Way's Star Formation Rate} \label{sec:sfr}
\subsection{The SFR Data} \label{sec:sfr_data}
Recent work by \citet*{Chomiuk} provides a thorough review and renormalization of Galactic star formation rate (SFR), \sfr, estimates made in the last three decades. Examining these original works reveals a discouraging scatter in the derived results, spanning the range of 1 to 10 \sfrunits. This proves to be predominantly due to a heterogeneous mixture of initial mass functions (IMFs) and stellar population synthesis (SPS) models used. The authors translate these results all to a uniform choice of the Kroupa broken-power-law IMF \citep{Kroupa} as well as the Starburst99 v5.1 SPS model \citep{Starburst}. As a result, these studies, which encompass many different methods and observational surveys, all collectively are in general agreement with each other after being placed on an equal footing, converging to a SFR of $\sfr=1.9\pm0.4$ \sfrunits.  We refer the reader to Table 1 of \citet*{Chomiuk} for the data we use to estimate the Milky Way SFR, as well as to their \S3 for a detailed discussion of the measurements. We do not utilize the first two entries in that table, the measurements from Smith et al. (1978) and G\"{u}sten \& Mezger (1982), as these were both updated by the Mezger (1987) radio free-free result.  Additionally, the error estimate on the Misiriotis et al. (2006) dust-heating measurement has been increased to 0.95 \massunits; i.e., due to the lack of error estimates we assign 50\% uncertainty to it.  This particular dataset is denoted $\D_{\text{SFR}}$ hereafter.

\subsection{Setting a Prior on \textbf{\sfr}} \label{sec:sfr_prior}

To place a prior on its SFR, we utilize the fact that the Milky Way is confidently known to be a spiral galaxy. In addition, it appears that the Galaxy has experienced a relatively quiet merger history, undergoing predominantly secular evolution with no significant interactions to spark a large burst of new stars \citep[e.g.,][]{Unavane,Mutch}.  Therefore, we agnostically assume that the Galactic SFR could be anywhere from 0 to 6 solar masses per year (before considering the data in \S\ref{sec:sfr_data}).  This is represented using a uniform probability distribution, such that

\begin{equation}
	P(\sfr)=
		\begin{cases}
		\frac{1}{6},& \text{if } 0 \leq \sfr \leq 6 \,\sfrunits \\
		0,		& \text{otherwise.}\\
		\end{cases}
\end{equation}

\subsection{SFR Results} \label{sec:sfr_results}
\begin{deluxetable*}{lcccccc}
\tablewidth{\textwidth}
\tablecaption{Combined SFR Results For Various Model Assumptions}
\tablehead{\multirow{2}{*}{Model} & Optimal & Combined \sfr$\pm1\sigma$ & \multirow{2}{*}{$k$\tablenotemark{b}} & \multirow{2}{*}{$\Delta$AIC} & \multirow{2}{*}{$\Delta$BIC} & \multirow{2}{*}{$\Delta\log_{10} K$} \\
& Value\tablenotemark{a} & (\sfrunits) & & & &}
\startdata
\\[-1.5ex]
\multicolumn{7}{c}{\underline{$f_{\text{good}}$ free --- Some of the measurements have inaccurate error bars.}} \\[1.5ex]
free-$n$ & $1.00$ & $1.65\pm0.20$ & 3 & 4.0 & 4.4 & -0.73 \\
free-$Q$ & $0.00$ & $1.66\pm0.21$ & 3 & 4.0 & 4.4 & -0.31 \\
free-$F$ & $0.26$ & $1.67\pm0.22$ & 3 & 4.0 & 4.4 & -0.25 \\
$P_\text{bad}$-flat & N/A & $1.65\pm0.20$ & 2 & 2.0 & 2.2 & -0.73 \\[1.5ex]
\multicolumn{7}{c}{\underline{$f_{\text{good}}=0$ --- All of the measurements have inaccurate error bars.}} \\[1.5ex]
free-$n$ & $1.00$ & $1.65\pm0.22$ & 2 & 2.0 & 2.2 & -1.59 \\
free-$Q$ & $0.00$ & $1.67\pm0.23$ & 2 & 2.0 & 2.2 & -0.58 \\
free-$F$ & \phn$0.16$\tablenotemark{c} & $1.69\pm0.26$ & 2 & 2.0 & 2.2 & -0.49 \\[1.5ex]
\multicolumn{7}{c}{\underline{$f_{\text{good}}=1$ --- None of the measurements have inaccurate error bars.}} \\[1.5ex]
all-``good''\tablenotemark{d} & N/A & $1.65\pm0.19$ & 1 & --- & ---  & --- 
\enddata
\tablenotetext{a}{The value of $n$, $Q$, or $F$ marking the peak of marginalize posterior PDF for these quantities in each model.}
\tablenotetext{b}{The number of free parameters in the model.  See \S\ref{sec:better_model}.}
\tablenotetext{c}{This corresponds to the lowest allowed value of $F$ in the model, $F^\text{MIN}$, where the smallest allowed error on any estimate in $\D_\text{SFR}$ is just the minimum error estimate; that is, $\sigma_{F,i}=F^\text{MIN}\mu_i^\text{MED}=\sigma_i^\text{MIN}$.  Since $F^\text{MIN}$ affects no estimates in $\D_\text{SFR}$, this is equivalent to setting $f_\text{good}$ to unity (see Equation \eqref{eq:likelihood_F}).  In fact, the HB analysis most strongly supports making no adjustment to the nominal error bars (i.e., the all-``good'' model where we always set $f_\text{good}=1$).}
\tablenotetext{d}{Since we use flat priors that are much broader the the likelihood PDF, this is equivalent to the inverse-variance weighted mean.}
\label{table:sfr}
\end{deluxetable*}
\begin{figure}
\centering
\includegraphics[trim=0.6in 1.25in 0.7in 0.5in, clip=true, width=1.0\columnwidth]{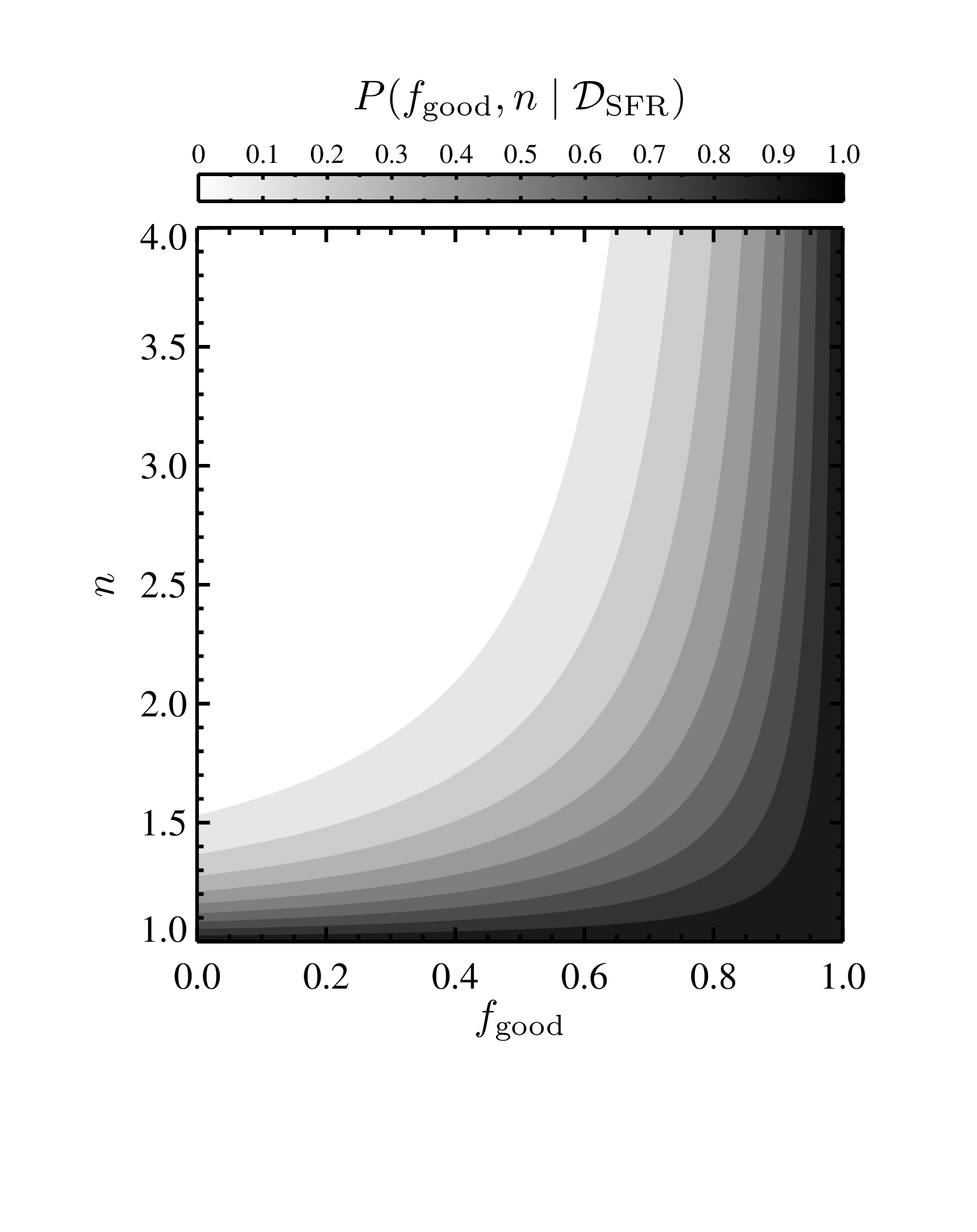}
\caption{Joint posterior probability distribution function (PDF), $P({f_{\text{good}}, n \mid \D_{\text{SFR}}})$, describing the probability of each possible value of the fraction of ``good'' estimates, $f_{\text{good}}$, (i.e., ones with accurate error bars) included in our Milky Way star formation rate (SFR) dataset, $\D_{\text{SFR}}$, and the scale factor $n$ needed to expand the error bars for any measurement treated as not ``good''.  This 2-dimensional probability distribution is produced from a hierarchical Bayesian (HB) analysis, where we have marginalized over all possible true values for the Milky Way's SFR.  For ease of reading, we have normalized the peak value to 1.  This plot clearly shows how strongly the HB analysis favors a model with minimally adjusted error bars on the individual estimates included in our analysis: the probability given the observed dataset is maximized if either $n$ is $\sim$1, and/or $f_{\text{good}}$ is $\sim$1.}
\label{fig:post_fg_n}
\end{figure}
\begin{figure}
\centering
\includegraphics[trim=0.7in 1.25in 0.7in 1.45in, clip=true, width=1.0\columnwidth]{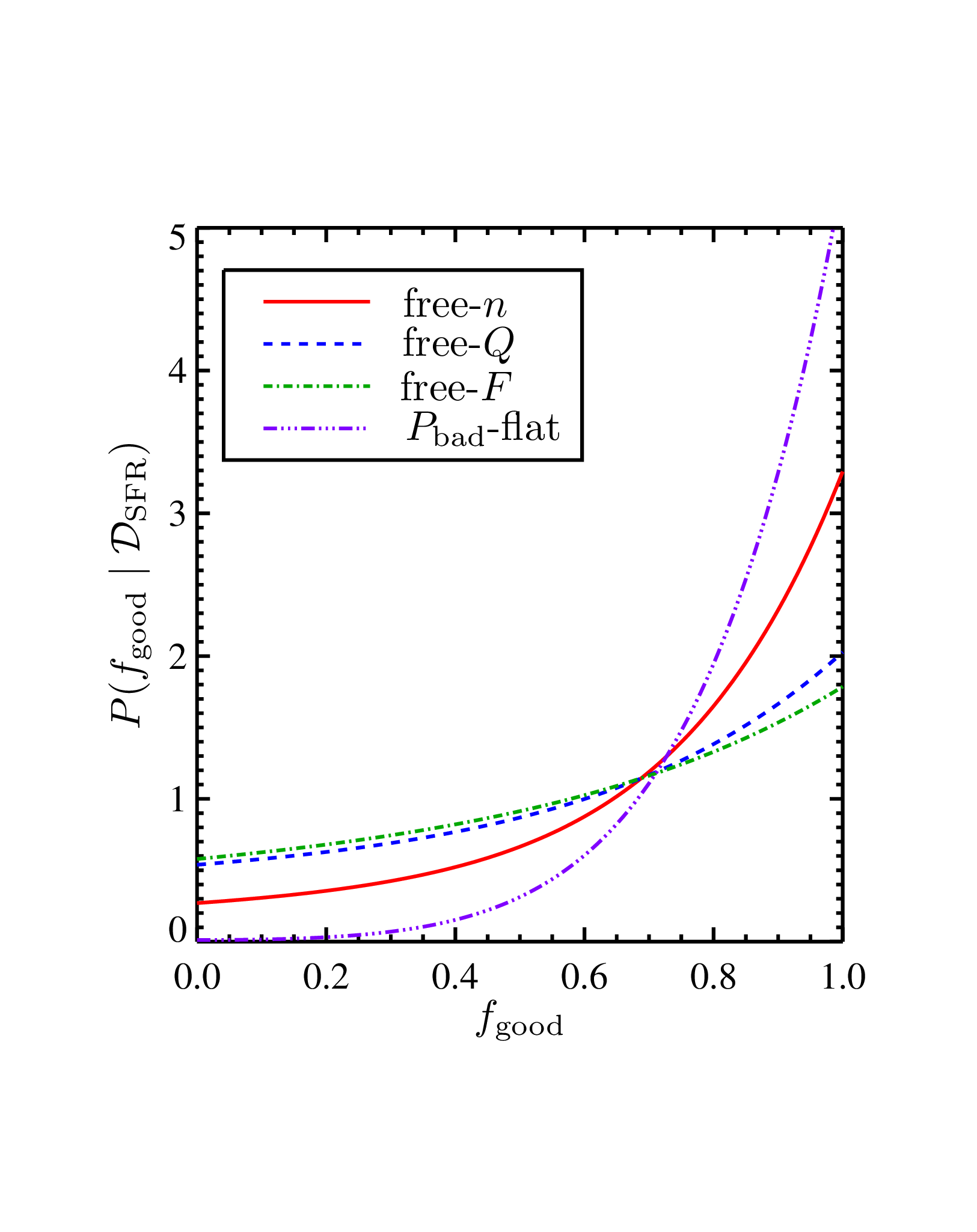}
\caption{Posterior PDF, $P({f_{\text{good}} \mid \D_{\text{SFR}}})$, for the fraction of ``good'' estimates (i.e., ones with accurate error bars), $f_{\text{good}}$, included in our Milky Way star formation rate (SFR) dataset, $\D_{\text{SFR}}$, for each model of ``bad'' measurements we consider.  Each curve is produced from a hierarchical Bayesian analysis, where we have marginalized over all other free parameters in the model.  Each corresponds to a different model of how to remedy the inclusion of ``bad'' measurements included in $\D_{\text{SFR}}$: the solid red curve corresponds to a case where we must multiply the error bars for ``bad'' measurements by a free parameter $n$; the dashed blue curve is produced by adding a fractional amount (given by the free parameter $Q$) of the median value of all estimates included in $\D_{\text{SFR}}$ in quadrature to their nominal error bars; the green dash-dotted curve results from imposing a floor on ``bad'' measurements' nominal error bars equal to a fraction amount (given by the free parameter $F$) of the median estimated SFR; and the purple triple-dot-dashed curve corresponds to a case where we model ``bad'' measurements as entirely wrong, and replace them with uniform probability distributions over the entire parameter space.  We see that all models peak at $f_{\text{good}}=1$, indicating that there is minimal tension between the measurements included in $\D_{\text{SFR}}$.  Smaller values of $f_{\text{good}}$ can be consistent with the data for most models, but only if $Q$ or $F$ is $\sim$0 or $n\sim1$, corresponding to no difference between ``good'' and ``bad'' measurements.}
\label{fig:post_fg_comp}
\end{figure}

Table \ref{table:sfr} shows the hierarchical Bayesian constraints on \sfr\ from each model that we consider.  The overwhelming similarity between the posterior results from these different models to a simple weighted average (corresponding to the last line in the table) suggests that nominal errors on each measurement in the SFR dataset are likely very accurate, if not overestimated.  In Figure \ref{fig:post_fg_n}, we show the joint posterior PDF for $f_{\text{good}}$ and $n$ (normalized to a peak value of 1); the posterior obtained by marginalizing over \sfr\ is maximized where $f_{\text{good}}$ and $n$ each approach unity.  Figure \ref{fig:post_fg_comp} shows the marginalized posterior probability for $f_{\text{good}}$ for each model which allows it to be a free parameter.  It is clear from the plot that modifying the error bars of "bad" measurements in any way yields little preference for a particular value of $f_{\text{good}}$, whereas throwing out ``bad'' estimates favors values near one.  This occurs because for values of $n$ near 1 (or $Q$ or $F$ near 0), $f_{\text{good}}$ has very little effect, so a broad range of $f_{\text{good}}$ values yield similar results.

\begin{figure}
\centering
\includegraphics[trim=0.45in 1.25in 0.8in 1.45in, clip=true, width=1.0\columnwidth]{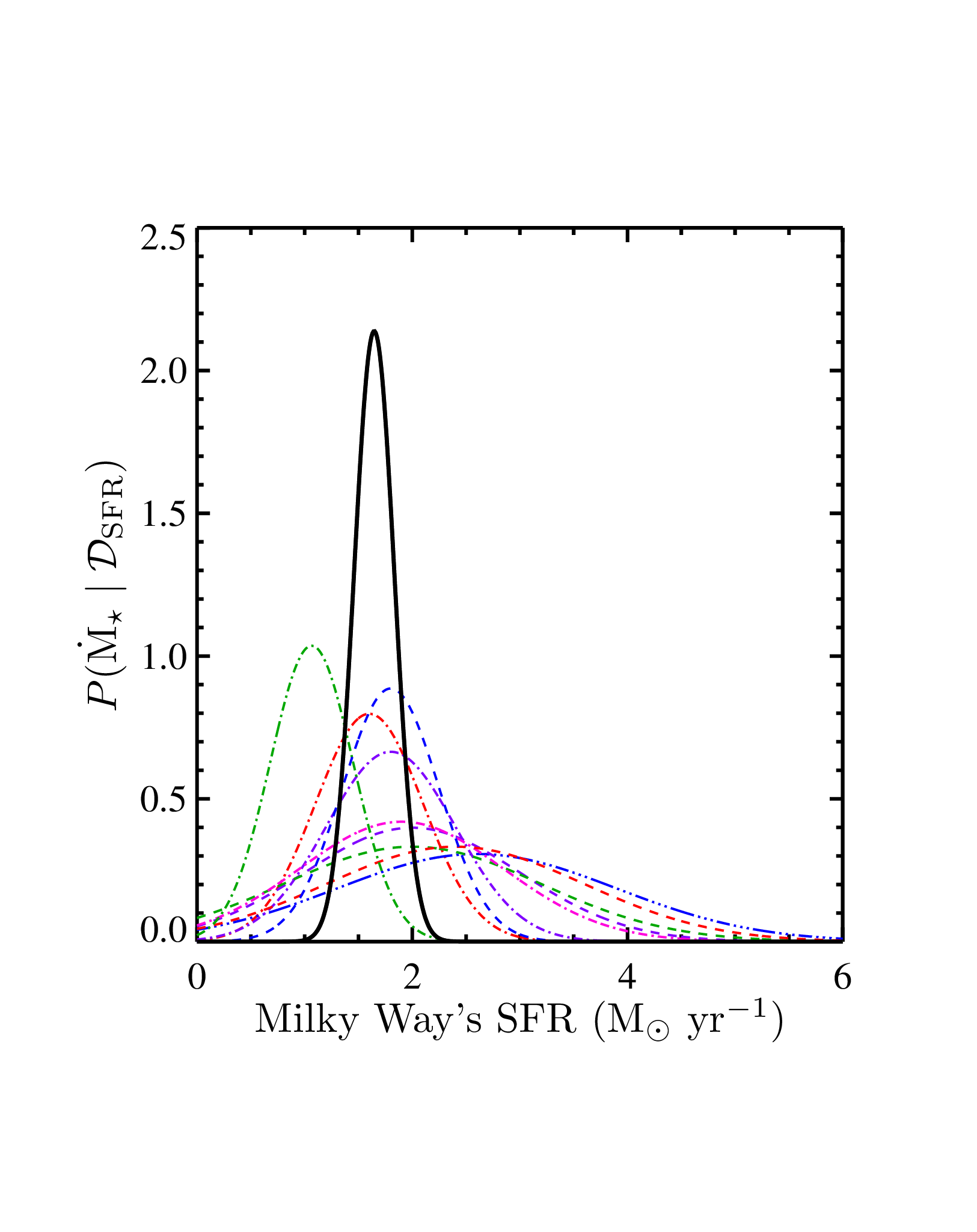}
\caption{Posterior PDF, $P({\sfr \mid \D_{\text{SFR}}})$, for the Milky Way's star formation rate (SFR) is shown as a solid black line.  This result is produced from the hierarchical Bayesian (HB) analysis where we assume all measurements have accurately estimated error bars; since we have assume flat priors over the entire parameter space, this is equivalent to the inverse-variance weighted mean.  This model is the simplest one which provides a good fit to the data, and hence is favored by both the Bayesian evidence and information criteria.  For comparison, we overlay the individual estimates from our SFR dataset as dashed/dotted colored lines.  We see that our methods yield a more tightly constrained estimate of the SFR of the Milky Way, while also being consistent with each individual estimate incorporated into the combined result.}
\label{fig:sfr_comp}
\end{figure}

Similar analyses to that in Figure \ref{fig:post_fg_n} for each of the models of ``bad'' measurements yields the same overall message: the measurements are sufficiently consistent with each other that little, if any, adjustment of the nominal errors is demanded.  In fact, we find that for any model for how to remedy ``bad'' estimates within the HB formalism, the data always drives towards a point in $k$-dimensional parameter space with few points treated as discrepant; i.e., $f_{\text{good}}\approx1$, and $\sfr\approx1.66$ \sfrunits; or, alternatively, $f_{\text{good}}<1$ but $n=1$ or $Q$ or $F=0$, which has equivalent effect.  The best model of the data (i.e., the one with largest Bayes Factor, $K$) turns out to be where we set $f_{\text{good}} = 1$, uniformly treating every estimate as ``good'' and hence requiring the fewest free parameters in the fit (only \sfr\ itself).  This yields an aggregate SFR for the Milky Way of $\sfr=1.65\pm0.19$ \sfrunits, which we choose as our fiducial result.  For comparison, we overlay the individual measurements in the SFR dataset on our fiducial hierarchical result in Figure \ref{fig:sfr_comp}.  We note that, as we are using flat priors, this scenario gives the same result as the inverse-variance weighted mean (IVWM).  The finding that the data are sufficiently compatible that a simple model is sufficient matches well with what we would conclude using the chi-squared statistic to judge goodness-of-fit.  Specifically, the IVWM for our sample of SFR measurements yields $\chi^2=3.83$, and we would expect a 95\% chance of this statistic falling in the range $[2.18,17.53]$.

\citeauthor{Chomiuk} present arguments suggesting that the Robitaille \& Whitney (2010) measurement may more accurately be treated as a lower limit rather than a best-fit value for the SFR.  Given that uncertainty, we investigate the effect of doubling the nominal errors in that measurement within our calculation.  Making this change yields a $<1\sigma$ shift in the mean of the posterior distribution. With this deweighting, the inclusion or exclusion of this measurement has little impact on our results (changing the consensus Milky Way SFR by no more than 0.17 \sfrunits).  A second concern arises from that fact that Equations \eqref{eq:likelihood_n}--\eqref{eq:likelihood_pbad_flat} are contingent on the individual measurements being independent of one another.  We note that, even though Bennett et al. (1994) and McKee \& Williams (1997) utilize the same \cobe\ data, the errors in their SFR estimates appear to be dominated by the differences between the assumptions made by the authors, and so we expect that we can treat them as independent measurements.  We show how our aggregate SFR estimate varies under these different treatments of the Robitaille \& Whitney (2010) measurement, as well as when excluding the Bennett et al. (1994) measurement, in Table \ref{table:sfr2}.

If all of the star formation rate estimates employed suffer from a common systematic error, this would not be reflected in the hierarchical Bayesian result.  For instance, if the Kroupa IMF is unlike the actual initial mass function in the Milky Way, our SFR estimates could all be systematically off from the true value in a similar way.  However, this error would cancel out when the Galaxy is compared to extragalactic objects, for which SFRs and stellar masses are generally calculated assuming Kroupa-like IMFs.  Similarly, \citeauthor{Chomiuk} have recalibrated each of the Milky Way \sfr\ estimates to the \citet{Kennicutt} assumption that a SFR of 1 \sfrunits\ produces a Lyman continuum photon rate of $9.26\times10^{52}$ photon s$^{-1}$ for a Salpeter IMF (this relationship is then reduced by a factor of 1.44 to convert it to the Kroupa IMF).  If this assumption is in error, then all the SFRs in our dataset would be affected.  However, once more, extragalactic SFRs would be off by the same factor, so that this systematic will drop out when comparing the Milky Way to other galaxies.

Apart from the IMF and the Lyman continuum rate to SFR relationship, common-mode systematics amongst all the SFR measurements appear to be unlikely.  This is due to both the diversity of techniques used to estimate the Galactic star formation rate and the wide range of assumptions made by the different studies that utilize the same basic techniques.  However, one might still worry that there are common underlying assumptions that may systematically offset SFR results from one method in comparison to all the others.  For instance, two of the nine studies we use estimate the Galactic SFR by modeling the population of young stellar objects (YSOs) found from infrared (IR) surveys of the Galaxy.  While these studies employ entirely different IR data, they are both contingent on assumptions about the properties of YSOs, which do not affect the other measurements, and in turn can systematically shift these results in unison with respect to the measurements utilizing different methods.  Our HB analysis assumes that the errors from each study are random compared to each other and hence does not address this type of common-mode effect (much as the inverse-variance weighted mean would not).

To test whether method-to-method variations are significant, we have performed bootstrap resampling of the SFR data by randomly drawing only one measurement out of those utilizing each measurement technique (e.g., all those based on the measured ionization rate or YSO counts).  Hence, each resampled dataset comprises four SFR estimates, each obtained from a unique measurement method, which we then use to calculate the HB posterior using the all-``good'' model (allowing no extra compensation for systematic errors).  We repeat this process 1,000 times, each time measuring the mean from the posterior distribution.  We then do a similar analysis, but where we instead draw four measurements at random from the entire dataset each time.  We find the standard deviation of the mean to be 0.22 if we use four measurements from different methods and 0.25 if we select four measurements completely at random.  This indicates that inter-method variations are negligible compared to intra-method random variations -- if anything, SFR measurements from {\it different} methods are more similar to each other than those which utilize the {\it same} technique.  We therefore can safely conclude that common-mode systematics do not have a large impact.

\section{The Mass of the Milky Way's Stellar Components} \label{sec:mstar}
In this section we describe the methods we use to produce improved estimates of the total stellar mass, \mt, in the Milky Way and its components.  To do so, we first make independent estimates of the stellar mass contained in the disk and bulge+bar components, \md\ and \mb\ respectively.  For \md, we assume the single-exponential model of the Galactic disk from \citet{Bovy2013}, and use Monte Carlo (MC) simulations to incorporate updated estimates of the Sun's Galactocentric radius, $R_0$.  We constrain \mb\ using our HB formalism, similar to the analysis done in \S\ref{sec:sfr} for \sfr, but now using the MC simulations to propagate uncertainties in the value of $R_0$ into the bulge mass posterior.  Lastly, we combine self-consistent realizations of \mb\ and \md\ to yield a probability distribution describing the total stellar mass.

\subsection{The Stellar Disk Model} \label{sec:mstar_diskmodel}

To model the structure of the Milky Way, we assume the stellar material of the disk is distributed in a single-exponential surface density profile.  Integrating this profile over all radii yields the total stellar mass,
\begin{equation}
\md = 2\pi\Sigma_{\star}(R_0)L^2_d\exp(R_0/L_d),
\label{eq:md}
\end{equation}
where $\Sigma_{\star}(R_0)$ is the surface density of stellar mass in the local neighborhood and $L_d$ is the exponential scale length of the disk.  Specifically, we constrain these parameters to be consistent with the measurements made by \citet[][see the Appendix for a discussion of alternative disk models]{Bovy2013}.
\begin{deluxetable}{llc}
\tablewidth{\columnwidth}
\tablecaption{Combined SFR Results For Various Data Assumptions}
\tablehead{\multirow{2}{*}{Model} & \multirow{2}{*}{Treatment of Data} & Combined \sfr$\pm1\sigma$ \\
& &  (\sfrunits)}
\startdata
all-``good'' & Including RW10\tablenotemark{a} with & $1.65\pm0.19$ \\
 & nominal errors (Fiducial) &  \\
all-``good'' & Including RW10 with & $1.77\pm0.21$ \\
 & errors doubled &  \\
all-``good'' & Excluding RW10 & $1.82\pm0.21$ \\
 & from the calculation &  \\
all-``good'' & Excluding B94\tablenotemark{b} & $1.63\pm0.19$ \\
 & from the calculation &
\enddata
\tablenotetext{a}{Robitaille \& Whitney (2010).}
\tablenotetext{b}{Bennett et al. (1994).}
\label{table:sfr2}
\end{deluxetable}

Using SDSS/SEGUE spectroscopic measurements, these authors have segregated a uniform sample of $\sim$16,000 G-type dwarf stars into 43 mono-abundance populations (MAPs) based on their position in [$\alpha$/Fe]-[Fe/H] space.  G-type dwarfs were considered to be the optimal tracers of the structure of the disk as they are most luminous stars whose main-sequence lifetimes are larger than the age of the disk at practically all metallicities.  Separated in this way, the spatial distribution of each MAP is well fit by a single-exponential profile both radially from the Galactic center and perpendicularly from the plane of the disk, indicating that the disk is likely composed of a more continuous distribution of populations rather than just the classical separation into thin and thick disk components \citep[see also][]{Bovy2012,Rix}.

By independently fitting an action-based distribution function and Galactic potential to each MAP in position-velocity phase-space, the authors are able to measure the vertical force at $|Z|\approx1$ kpc as a function of Galactocentric radius in the region $4\lesssim R\lesssim 9$ kpc; this quantity is directly proportional to the surface density of stellar mass, $\Sigma_{\star}$.  Including the contribution from all MAPs, the authors are able to make the first dynamical determination of the total stellar surface-mass density at the Galactocentric radius of the Sun, $\Sigma_{\star}(R_0)=38\pm4$ \massunits\ pc$^{-2}$; this is measured directly from the mass distribution of substellar objects, main-sequence stars, and stellar remnants, as opposed to inferring it from their luminosity \citep[cf.][]{Juric}.  Similarly, the mass-weighted scale length determined from all MAPs is $L_d=2.15\pm0.14$ kpc.

Assuming $R_0=8$ kpc and accounting for the uncertainty and covariance in $L_d$ and $\Sigma_{\star}(R_0)$, \citet{Bovy2013} find $\md=4.6\pm0.3\times10^{10}$ \massunits.  They also find that increasing $R_0$ to 8.5 kpc in their model produces a $1.5\times10^{10}$ \massunits\ increase in \mt; this effect approximately scales linearly for intermediate radii.  The scale length and local surface density of the disk are independent of $R_0$.  For our purposes, it is important to include the current uncertainties in $R_0$ into our model of the Milky Way, and so we employ the following process for this study:
\\ \\
\textbf{\textrm{I})} We model the covariance between the local surface density and scale length in this model by calculating $\Sigma_{\star}(R_0)$ as a function of $L_d$.  To do so, we first fit for a linear mapping between these two parameters based on the 2D joint-posterior PDF that describes them \citep[provided by][priv. comm.]{BovyPriv}.  This relation gives the appropriate value of $\Sigma_{\star}(R_0)$ for a given value of $L_d$.  We then determine what uncertainty in $\Sigma_{\star}(R_0)$ would yield the \citet{Bovy2013} error of $0.3\times10^{10}$ \massunits\ in \md\ using Equation \eqref{eq:md} after this covariance and the errors in $L_d$ are accounted for.
\\ \\
\textbf{\textrm{II})} We assume $R_0=8.33\pm0.35$ kpc from \citet{Gillessen}, which is estimated by fitting the orbital parameters of 28 stars in near-orbit of the Galaxy's central massive black hole, building upon 16 years of observations, and taking into account both random and systematic errors.  This result is in excellent agreement with other recent measurements of $R_0$ \citep[e.g.,][]{Ghez, Vanhollebeke, Sato, Chatzopoulos}, with a large enough error to encompass the variation in results amongst different methods.
\\ \\
\textbf{\textrm{III})} Given $\textrm{d}\md/\textrm{d}R_0$ from \citeauthor{Bovy2013}, we can then predict both the mass that would have been measured by \citeauthor{Bovy2013} for different $R_0$ and the uncertainty in that estimate due to measurement errors in $L_d$ and $\Sigma_{\star}(R_0)$ alone.
\\ \\

The overall goal of this study is to produce updated and accurate estimates of the total stellar mass and star formation rate of the Milky Way that may be directly compared with those properties measured for any external galaxy.  In particular, we aim to be consistent with the definition of \mt\ and \sfr\ as measured for external galaxies in the MPA-JHU catalogs \citep{Brinchmann2004}, which assume the Kroupa IMF and that \mt\ includes the contribution from both main-sequence stars and remnants, but not brown dwarfs (BDs), in accordance with the assumptions made for the stellar spectral evolution models of \citet{BC03}.  The dynamical estimate of $\Sigma_{\star}(R_0)$ from \citet{Bovy2013} effectively includes BDs, so one way to get the density we want is to subtract them off.

The BD mass function (i.e., $\xi(M)\propto M^{-\alpha}$ where $0.005<M/\massunits<0.1$) appears to have power-law index in the range $-0.5\lesssim\alpha\lesssim0.5$ \citep{Cruz,Kirkpatrick,DayJones,Burningham}.  We normalize the mass function so that the portion corresponding to L0--L3 BDs, i.e., objects with masses in the range $0.03<M/\massunits<0.05$ \citep[following the models of][updated 1998]{D'Antona}, integrates to a total number density of $1.7\times10^{-3}$ pc$^{-3}$ \citep[matching][]{Cruz}.  Accounting for the range of possible $\alpha$ values, we then integrate $\xi(M)\times M$ over the entire BD mass range to find a mass density of $\sim$2.5--$3\times10^{-4}$ \massunits\ pc$^{-3}$.  Lastly, as required for a single-exponential disk profile, we multiply this by 2 times a scale height of the disk of $h_z=400$ pc \citep[][]{Bovy2013}, yielding predicted BD surface densities of $\sim$5--6 \massunits\ pc$^{-2}$.  Earlier measurements, however, have yielded lower surface densities at $\sim$0.5--2 \massunits\ pc$^{-2}$ \cite[][and references therein]{Fuchs,Flynn}.  Based on these analyses, we conservatively assume that the local surface density of BDs is in the range $0.5\lesssim\Sigma_{\text{BD}}(R_0)\lesssim6$ \massunits\ pc$^{-2}$; assuming a uniform distribution over this range, we find $\Sigma_{\text{BD}}(R_0)=3.25\pm1.59$ \massunits\ pc$^{-2}$.  Subtracting this from \citeauthor{Bovy2013}'s dynamical estimate, we find $\Sigma_\star(R_0) = 34.75\pm4.30$ \massunits\ pc$^{-2}$.

Alternatively, we could compare to the stellar surface density estimate from \citet{Bovy2012}, which omits BDs but also white dwarfs (WDs) that we want to include.  To remedy this, we collect a number of literature estimates of the spatial density of local white dwarfs: \citet{Oswalt} find 0.0076 pc$^{-3}$; \citet{Leggett} find $0.0034$ pc$^{-3}$; \citet{Jahrei} find 0.005 pc$^{-3}$; \citet{Knox} find 0.00416 pc$^{-3}$; \citet{Holberg} find 0.0048 pc$^{-3}$; and \citet{Sion} find 0.0049 pc$^{-3}$.  Using the mean and standard deviation of this sample, we assume the local volume density of WDs is $\rho_\text{WD}=5.0\pm1.4\times10^{-3}$ pc$^{-3}$.  Following our disk model, we multiply this by 2 times a scale height of $h_z=400$ pc \citep[][and ascribing 10\% error]{Bovy2013}  and an average WD mass of $\langle M\rangle_\text{WD}=0.65\pm0.01$ \massunits\ \citep{Falcon} to find the local surface density of white dwarfs $\Sigma_\text{WD}(R_0)=\rho_\text{WD}\times 2h_z\times\langle M\rangle_\text{WD}=2.6\pm0.8$ \massunits\ pc$^{-2}$.  As a consistency check, we add this to the \citet{Bovy2012} photometric estimate of $32.0\pm3.2$ \massunits\ pc$^{-2}$ (due to main-sequence stars, assuming the Kroupa IMF and 10\% uncertainty), yielding $\Sigma_\star(R_0) = 34.6\pm3.3$ \massunits\ pc$^{-2}$; this is in excellent agreement with the corrected dynamical estimate, which we adopt for the remainder of this study, given its more conservative error estimate.  For convenience, we tabulate all parameters of our disk model, and their interdependencies, in Table \ref{table:params}.  Plugging these expressions into Equation \eqref{eq:md}, any realization of the total stellar disk mass from our Monte Carlo simulations is a function of the Galactocentric radius of the Sun and the scale length (independently drawn from their attendant probability distributions):
\begin{align}
&\md(R_0,L_d) = 2\pi(31.75L_d/\textrm{kpc}-33.5125\pm2.89)\nonumber\\
&\times(L_d/\text{kpc})^2\exp(8\,\text{kpc}/L_d)+3(R_0/\text{kpc}-8)\times10^{10}\,\massunits.
\label{eq:md_full}
\end{align}
\begin{deluxetable*}{lllll}
\tablewidth{0pt}
\tablecaption{Disk Model Parameters}
\tablehead{Parameter & Value & Units & Description & Reference}
\startdata
$R_0$ & $8.33\pm0.35$ & kpc & Galactocentric radius of the Sun & \citet{Gillessen} \\
$L_d$ & $2.15\pm0.14$ & kpc & Scale length of exponential disk & \citet{Bovy2013} \\
\multirow{2}{*}{$\Sigma_\star(R_0)$} & \multirow{2}{*}{$31.75L_d/\textrm{kpc}-33.5125\pm2.89$} & \multirow{2}{*}{\massunits\ pc$^{-2}$} & Local surface mass density of MS & \multirow{2}{*}{\citet[][priv. comm.]{BovyPriv}} \\
 & & & stars and remnants for a given $L_d$\tablenotemark{a} & \\
\multirow{2}{*}{$\md(R_0,L_d,\Sigma_\star(R_0))$} & \multirow{2}{*}{See Equation \eqref{eq:md_full}} & \multirow{2}{*}{\massunits} & The total stellar disk mass &  \multirow{2}{*}{\citet{Bovy2013}} \\
& & & from Equation \eqref{eq:md}, given $R_0$ and $L_d$ &
\enddata
\tablenotetext{a}{See \S\ref{sec:mstar_diskmodel}.  This relationship accounts for the covariance between the dynamical estimates of $L_d$ and $\Sigma_\star(R_0)$ from \citet{Bovy2013}, while also subtracting the contribution from brown dwarfs, $\Sigma_{\text{BD}}(R_0)$.}
\label{table:params}
\end{deluxetable*}

\subsection{Monte Carlo Techniques} \label{sec:monte_carlo}
As we shall see in \S\ref{sec:mstar_data}, many estimates included in our Galactic bulge+bar mass dataset are dependent on assumptions made about the values of $R_0$ and/or \md, and hence depend upon all of our disk model parameters.  However, as denoted in Table \ref{table:params}, the true value of these parameters are not perfectly known; we can describe our prior knowledge of each one by a probability density function.  In order to propagate the uncertainties in these parameters' values into the resulting posteriors describing the stellar mass of the disk and bulge, as well as their combination for the total stellar mass of the Galaxy, we perform a series of Monte Carlo (MC) simulations.

Unlike the parameters describing our ``bad''-measurement models ($\Theta$), we are uninterested in producing new results for our disk model parameters.  We have much greater confidence in the priors we have chosen for $R_0$, $L_d$, and $\Sigma_\star(R_0)$ based on the more direct observations described in the previous section than anything we would infer about those quantities from the set of bulge mass measurements.  Therefore, in this application we want to ensure that the results of our HB formalism will reflect the information contained in the priors in Table \ref{table:params}.  We will collectively denote the set of disk parameters as \N, as we are uninterested in allowing the \mb\ data to modify any conclusions about what the values of those ``nuisance'' parameters may be.  Formally, we are making the assumption that $P({\N \mid \D})=P(\N)$ --  i.e., that our state of knowledge of the nuisance parameters is not affected by the bulge dataset -- and we refer to this as the ``strict-prior'' method below.  In this framework, our procedure is as follows:
\\ \\
\textbf{\textrm{I})} We independently and randomly generate a sample of 10$^3$ values of $R_0$ and $L_d$, which collectively reproduce the respective mean and standard deviation given in Table \ref{table:params}.  Then for each realization of the scale length of the disk, $L_{d,i}$, we randomly draw a corresponding value of $\Sigma_\star(R_0)_i$ based on the relationship listed in Table \ref{table:params}.
\\ \\
\textbf{\textrm{II})} For each of the 10$^3$ sets of randomly drawn parameters, $\N_i=\left\{R_0,L_d,\Sigma_\star(R_0)\right\}_i$, we calculate the corresponding disk mass, \mdi, using Equation \eqref{eq:md_full}.  The result is a distribution of possible Galactic disk masses, which we normalize to produce $P({\md \mid \D})$.  Note that this is consistent with our strict-prior assumption that $P({\N \mid \D})=P(\N)$, and so the fraction of times that \mdi\ calculated from the $\N_i$ occurs in the Monte Carlo will be proportional to $P(\N)$.
\\ \\
\textbf{\textrm{III})} We determine the value of each literature \mb\ measurement (listed in Table \ref{table:mbulge}) that would have been obtained assuming these parameters are correct.  This ensures that all measurements make consistent assumptions about the structure of the Galaxy, allowing them to be combined fairly.  We refer the reader to the following section for these details.
\\ \\
\textbf{\textrm{IV})} For each iteration of \textbf{III}, along with the prior chosen for \mb, which we will detail in \S\ref{sec:mstar_prior}, we calculate the joint $k$-dimensional posterior $P({\Theta \mid \D, \N_i})$ via an HB analysis, as described in \S\ref{sec:bayes}.  After marginalizing out the other parameters in $\Theta$, each realization of the bulge mass posterior can be written
\begin{equation}
P({\mb \mid \D, \N_i}) = \frac{P({\D \mid \mb, \N_i})P(\mb)}{P({\D \mid \N_i})}, 
\label{eq:bayes_nuisance}
\end{equation}
applying BayesÕ theorem as usual.  By the definition of marginalization, $P({\mb \mid \D}) = \int P({\mb, \N \mid \D})\,\mathrm{d}\N$; and applying the definition of joint probability and our strict-prior assumption, this is equal to $\int P({\mb \mid \D, \N}) P(\N)\,\mathrm{d}\N$.  If we draw values $\N_i$ from our prior distribution $P(\N)$, this integral will be equal to 
\begin{equation} 
\displaystyle\lim\limits_{n\to\infty}\frac{1}{n}\displaystyle\sum\limits_{i=1}^{n} P({\mb \mid \D, \N_i}),
\end{equation}
since the fraction of times that $\N_i$ occurs in the Monte Carlo will be proportional to $P(\N)$.  Note also that Bayes' theorem allows us to rewrite the denominator in Equation \eqref{eq:bayes_nuisance} as $P({\D \mid \N_i})=P({\N_i \mid \D})P(\D)/P(\N_i)$, and by applying our strict-prior assumption this reduces to simply $P(\D)$.  Thus, in order to construct the combined \mb\ result for a given model of ``bad'' measurements, we are able to simply average the individual posteriors from the Monte Carlo realizations:
\begin{equation}
P({\mb \mid \D}) = \frac{1}{1000} \displaystyle\sum\limits_{i=1}^{1000}\frac{P({\D \mid \mb, \N_i})P(\mb)}{P({\D})}.
\label{eq:posterior_sum}
\end{equation}
The posterior PDF for each of the other parameters of the ``bad''-measurement model are calculated in the same way.  In addition, we record the AIC, BIC, and $K$ for each iteration, yielding a distribution of values.  In practice, we find that 10$^3$ realizations produces a standard error in the median $\Delta\log_{10} K$/AIC/BIC values much smaller than 0.5, which is sufficient to assess differences in the utility of different models securely (compared to our $\Delta=2$ criterion for significance).
\\ \\
\textbf{\textrm{V})} We also produce a posterior for the total stellar mass from each iteration, $P({\mt \mid \D, \N_i})$, in a model-consistent manner by simply defining $\mti=\mb+\mdi$.  Again, we can average the individual posteriors, $P({\mt \mid \D, \N_i})$, to obtain $P({\mt \mid \D})$, similar to Equation \eqref{eq:posterior_sum}.  Lastly, we also calculate the model-consistent posterior describing the bulge-to-total mass ratio by normalizing the distribution of values $(B/T)_i=\langle P({\mb \mid \D, \N_i})\rangle/\langle P({\mt \mid \D, \N_i})\rangle$ to integrate to unity, where angled brackets denote the mean of the enclosed posterior distribution. \\

One way of thinking about the strict-prior method is in terms of the Bayesian definition of probability as a degree of belief: $P({\mb \mid \D, \N_i})$ represents what we would conclude about \mb\ based on living in a Galaxy with parameters $\N_i$.  The prior $P(\N)$ represents our belief of how likely the parameter values $\N_i$ are to be correct compared to other possible values, and hence is the correct weighting to determine what we would believe about $P({\mb \mid \D})$, taking into account all possible values of \N\ and how probable we believe each of those values are.  This assumption is appropriate here, given that the bulge mass determinations depend only very indirectly on disk parameters, so we would place little faith in constraints on disk properties that came from the bulge mass dataset as opposed to the much more direct methods now available.

An alternative method would be to sum the \textit{likelihoods}, $P({\D \mid \mb, \N_i})$, and normalize that result to unity (which we will refer to as the ``weak-prior'' method).  This is equivalent to calculating the evidence-weighted average of the posteriors calculated in the course of the strict-prior method.  In this case, the posterior we calculate incorporates the assumption that $P({\N \mid \D})\neq P(\N)$: i.e., that our state of knowledge of the nuisance (disk) parameters should be influenced by the set of bulge mass measurements.  Hence, all the parameters in Table \ref{table:params} would be treated in the same way as those contained in $\Theta$, differing only in the informativeness and nature of the priors applied.  Effectively, the weak-prior method assumes that values of the parameters \N\ under which the data were most likely to have been observed should be given greatest weight, even if they are disfavored by our priors.

We have much greater confidence, however, in the priors we have chosen for $R_0$, $L_d$, and $\Sigma_\star(R_0)$ based on more direct observations than anything we would infer from the set of bulge mass measurements, and so the weak-prior method appears to be inappropriate here.  In practice, the posteriors for the disk mass, $P({\md \mid \D})$, and hence also for the total mass, $P({\mt \mid \D})$, differ significantly (with a mean differing by $\sim$1$\sigma$) between the two methods; however, the bulge mass estimate differs little between the strict- and weak-prior methods (with a mean differing by $\sim$0.1$\sigma$).

\subsection{A Uniform Sample of Bulge+Bar Mass Measurements} \label{sec:mstar_data}

In this study, we define \mb\ as the excess mass over a single-exponential disk in the total stellar mass budget for the Galaxy.  We begin by searching the literature for measurements of the combined stellar mass of the bulge, pseudo-bulge, and/or bar components of the Milky Way, which collectively fall into the category of \mb.  A diverse set of methods, models, and observations have been used to make these estimates.  For instance, \citet{Dwek95} photometrically determined the Galactic bulge morphology, luminosity, and mass using triaxial bar-like models constrained by the \cobe/\emph{DIRBE} observations, measuring $\mb=1.3\pm0.5\times10^{10}$ \massunits.  In contrast, \citet{KZS02} consider $\Lambda$CDM-based models for the Milky Way, accounting for the mass and angular momentum of the DM halo, constrained by a variety of kinematic measurements, to estimate that $\mb\approx1\times10^{10}$ \massunits.  Alternatively, \citet{Picaud04} use Monte Carlo techniques to fit the Besan{\c c}on model of stellar population synthesis to the observed near-IR luminosity of the bulge using observations from the DENIS survey, finding $\mb=2.4\pm0.6\times10^{10}$ \massunits.  These are only a few of the alternatives; we provide the entire list of studies included in our Galactic bulge+bar mass dataset, hereafter denoted as $\D_\text{M}$, in Table \ref{table:mbulge} for reference.

The incorporated studies use a heterogeneous mixture of assumptions and models for the Galaxy; we follow the basic model of \citet{Chomiuk} and attempt to place them on a uniform basis here.  One of the most common sources of variation is the adopted value of the Galactocentric radius of the Sun's orbit, $R_0$.  For instance, using the virial theorem one can demonstrate that kinematic estimates of the Galactic bulge+bar mass will be directly proportional to $R_0$.  On the other hand, photometric estimates of the luminosity of the bulge, based on the flux measured from our location, should scale as $R_0^2$.

A bulge mass inferred from the microlensing event rate observed toward the Galactic center, however, has a more complex dependence on $R_0$.  The measured rate of microlensing events can be directly converted into an optical depth, $\tau=\tau_\text{bulge}+\tau_\text{disk}$, which is the sum of the contribution from the stars in the bulge and the intervening part of the disk.  \citet{Hamadache06} show that $\mb\propto\tau_\text{bulge}\times R_\text{bulge}$, where $R_\text{bulge}$ is the radius of the bulge, which itself should be proportional to $R_0$ based on geometric arguments (i.e., $R_\text{bulge}=R_0\tan\theta/2$, where $\theta$ is the angular diameter of the bulge).  Unfortunately, the typical contribution from disk stars, which we would need to subtract in order to obtain $\tau_\text{bulge}$, is $\sim\tau/3$ \citep[][and references therein]{Sumi,Hamadache06} and turns out to be highly sensitive to the chosen value of $R_0$.  As an example, if we are to assume that $\tau_\text{disk}$ is proportional to the mass of disk stars between us and the bulge, which in turn depends on the total mass of the disk, the disk model described above yields $\text{d}\ln\mb/\text{d}\ln R_0\approx5.7$; i.e. $\tau_\text{disk}$ should scale roughly as $R_0^6$.  Based on this complexity, we assume for the purposes of this study that estimates of \mb\ based on measurements of $\tau$ do not scale will with $R_0$.

In order to combine the \mb\ estimates in our dataset into one aggregate result, after choosing a value of $R_0$ from our \citet{Gillessen} prior, we renormalize each result to that $R_0$ value using the appropriate scaling relation.  Unless otherwise noted, the central value and error bar are scaled by the same factor in order to ensure the fractional error remains unchanged (this is equivalent to holding the error bars constant in log space), as essentially none of the literature estimates include the uncertainty in $R_0$ in their error estimates (a problem that our Monte Carlo technique will fix).  For reference, we list the type of observational constraint and appropriate $R_0$ power-law index used to scale to each \mb\ estimate in Table \ref{table:mbulge}.  We note that if we assume that microlensing rate--based measurements of \mb\ scale as $R_0^1$ (as bulge stars are the dominant contribution to $\tau$), rather than treating them as independent of $R_0$, the change in our results proves to be negligible.
\begin{deluxetable*}{lccccc}
\tablewidth{\textwidth}
\tablecaption{The Galactic Bulge+Bar Mass Dataset}
\tablehead{\multirow{2}{*}{Reference} & $\mb\pm 1\sigma$ & $R_0$ assumed & \multirow{2}{*}{Constraint type}  &  \multirow{2}{*}{$\beta$\tablenotemark{a}} & $\mb\pm 1\sigma (R_0=8.33\text{kpc})$ \\
 & ($10^{10}$ \massunits) & (kpc) & & & ($10^{10}$ \massunits)}
\startdata
\citet{Kent92} & $1.69\pm0.85$ & $8.0$ & Dynamical & $1$ & $1.76\pm0.88$ \\
\citet{Dwek95} & $2.11\pm0.81$ & $8.5$ & Photometric & $2$ & $2.02\pm0.78$ \\
\citet{HanGould95} & $1.69\pm0.85$ & $8.0$ & Dynamical &$1$ & $1.76\pm0.88$ \\
\citet{Blum95} & $2.63\pm1.32$ & $8.0$ & Dynamical & $1$ & $2.74\pm1.37$ \\
\citet{Zhao96} & $2.07\pm1.03$ & $8.0$ & Dynamical & $1$ & $2.15\pm1.08$ \\
\citet{Bissantz97} & $0.81\pm0.22$ & $8.0$ & Microlensing & $0$ & $0.81\pm0.22$ \\
\citet{Freud98}\tablenotemark{b} & $0.48\pm0.65$ & \ldots & Photometric & \ldots & $0.48\pm0.65$ \\
\citet{Dehnen98} & $0.61\pm0.38$ & $8.0$ & Dynamical & $1/2$ & $0.62\pm0.38$ \\
\citet{Sevenster99} & $1.60\pm0.80$ & $8.0$ & Dynamical & $1$ & $1.66\pm0.83$ \\
\citet{KZS02} & $0.94\pm0.29$ & $8.0$ & Dynamical & $1$ & $0.98\pm0.31$ \\
\citet{Bissantz02}\tablenotemark{c} & $0.84\pm0.09$ & $8.0$ & Dynamical & $1$ & $0.87\pm0.09$ \\
\citet{HanGould03} & $1.20\pm0.60$ & $8.0$ & Microlensing & $0$ & $1.20\pm0.60$ \\
\citet{Picaud04} & $0.54\pm1.11$ & $8.5$ & Photometric & $0$ & $0.54\pm1.11$ \\
\citet{Hamadache06} & $0.62\pm0.31$ & None & Microlensing & $0$ & $0.62\pm0.31$ \\
\citet{Wyse06} & $1.00\pm0.50$ & None & Historical review & $0$ & $1.00\pm0.50$ \\
\citet{Lopez07} & $0.60\pm0.30$ & $8.0$ & Photometric & $2$ & $0.65\pm0.33$ \\
\citet{Calchi08} & $1.50\pm0.38$ & $8.0$ & Microlensing & $0$ & $1.50\pm0.38$ \\
\citet{Widrow08} & $0.90\pm0.11$ & $7.94$ & Dynamical & $1$ & $0.95\pm0.12$
\enddata
\tablecomments{\protect{B}ulge mass estimates, \mb, listed in this table have been converted to the Kroupa IMF.  See \S\ref{sec:mstar_data} for further notes on individual estimates.}
\tablenotetext{a}{$\beta$ denotes the assumed relationship between each \mb\ estimate, based on the constraint type, and the $R_0$ assumed; i.e. $\mb\propto R_0^\beta$.}
\tablenotetext{b}{Value (in both cases) calculated assuming $\md=5.17\times10^{10}$ \massunits\ and $R_0=8.33$ kpc.  $B/D$ results are published as a function of $R_0$, so no scaling relation needs to be assumed.}
\tablenotetext{c}{Values provided by \citet{McMillan}.}
\label{table:mbulge}
\end{deluxetable*}

As discussed in \S\ref{sec:mstar_diskmodel}, our definition of stellar mass includes the contribution from remnants, but not that from substellar objects.  Kinematically derived measurements of the bulge mass in our dataset will not reflect this distinction.  Therefore, we multiply the results from dynamical measurements by a normalization factor of $0.94\pm0.02$ to exclude the contribution from brown dwarfs.  We derive this scale factor by varying the power-law index of the IMF in the brown dwarf mass range ($0.005< M/\massunits<0.01$) from -0.5 to 0.5 \citep{Cruz,Kirkpatrick,DayJones,Burningham} and including their contribution within the \citet{BC03} model using solar metallicity and the Kroupa IMF over the main-sequence mass range.

Unlike dynamically constrained models of the Milky Way, which are indifferent to the mass distribution of the stellar populations, those which rely on photometric constraints (and thus mass-to-light ratios) to estimate \mb\ will depend strongly on the IMF of choice.  It is important then to set all measurements of this type on the same footing before applying the HB analysis by converting them to the same IMF.  As previously discussed, we choose that to be the Kroupa IMF in accordance with the MPA-JHU spectrum measurements \citep{Brinchmann2004}.

The HB analysis we apply to estimate \mb\ requires an error estimate for each independent measurement.  Many of the studies in the literature, however, do not provide error estimates.  We therefore must estimate the uncertainty in each measurement lacking an error bar in a uniform and unbiased manner; we do this in either one of two ways.  First, for any study that lists a single \mb\ result sans an error bar, we conservatively choose the error to be 50\% of the value.  However, if the authors instead provide a list of results corresponding to a variety of models or parameters explored, we then find the standard deviation of these listed values and add this in quadrature to 25\% of their favored value to produce an error estimate.

Lastly, as we will detail below, the central value of many of the \mb\ estimates in our dataset will vary with the disk parameters that we draw for each MC simulation.  In such a case, it becomes problematic to use the median estimated value (which will vary) as a reference value that multiplies $Q$ or $F$ in those models of unreliable data.  Consequentially, for the stellar mass portion of this study we always multiply $Q$ and $F$ by $10^{10}$ \massunits\ when augmenting the errors estimates of ``bad'' measurements.  For example, this means that $Q=0.5$ corresponds to adding $0.5\times10^{10}$ \massunits\ in quadrature to the nominal error bars when accounting for the possibility that a measurement is inaccurate (compare to Equation \eqref{eq:likelihood_Q}).

In the following list, we detail any studies included in our analysis of \mb\ that require further special treatment to be comparable to the others in Table \ref{table:mbulge}:
\\ \\
\textbf{\citet{Picaud04}:} This work uses the Besan{\c c}on model of stellar population synthesis \citep[see also][]{Robin03} to simultaneously constrain the mass of the bulge and thin disk through direct comparison with near-infrared star counts in $\sim$100 windows of low extinction in the Galaxy.  The thin disk is divided into 7 distinct age components with a two-slope IMF, while the bulge is modeled as a single older population with a Salpeter IMF; all of these populations are converted to follow the Kroupa IMF as the first step in our renormalization process.  Additionally, the Besan{\c c}on model features a double-exponential profile for the thin disk.  This disk model has a hole at the center, so the same amount of mass at the center of the Galaxy would correspond to a greater bulge mass than in a non-holed model.  For consistency, we integrate the holed density profile to determine the mass of the Besan{\c c}on thin disk within the bulge radius of 3.71 kpc, and subtract this value from the single-exponential disk mass within the same radius for each realization of our disk model.  The difference we calculate corresponds to the portion of the \citeauthor{Picaud04} bulge mass estimate ($2.4\pm0.6\times10^{10}$ \massunits) that has already been included in the Bovy disk, so we subtract it off.  Given that this correction is uncertain, we also add in quadrature to the nominal errors an amount equal to 50\% of the holed-disk correction when assuming $R_0=8.33$ kpc; i.e., 50\% of $1.86\times10^{10}$ \massunits.  \citeauthor{Picaud04} find that changes in the model mass when $R_0$ is changed are comparable in impact to the other variations amongst the models studied, and so we do not apply any $R_0$ scaling beyond this correction.  Changes in $R_0$ affect their models in multiple ways, rendering any simple scaling of their \mb\ value unsuitable.  In the end, the results of \citet{Picaud04} are included in our dataset as a value of $\mb=0.54\pm1.11\times10^{10}$ \massunits\ for $R_0=8.33$ kpc and best-fit Bovy disk parameters; for simplicity of analysis, this error estimate is used regardless of disk parameters.
\\ \\
\textbf{\citet{Zhao96}:} This work models the disk with the \citet[][hereafter NM]{Miyamoto} potential, whose profile roughly resembles that of the double-exponential disk.  Ideally, we would make the same correction using our disk model as we have done with \citet{Picaud04}.  However, the authors have normalized the NM profile to produce a total disk mass $\md=8\mb$, chosen so as to incorporate the dynamical effects of the dark halo and produce a flat rotation curve at $1<R<3$ kpc from the Galactic center.  We find, however, that this produces an extremely high surface density at nearly all radii within the disk, and indeed the authors note that this model does not fit the \cobe\ map data except near the center.  Since we cannot disentangle the dark mass from the stellar mass in this model, we do not attempt to do a correction as  for \citet{Picaud04}, since we would not have confidence in its accuracy.  To account for that inability to correct, we increase the errors on this estimate to 50\% of the derived bar mass.  Lastly, as this is a dynamical measurement, we remove the contribution of brown dwarfs to the bulge mass budget by multiplying by a factor of $0.94\pm0.02$.  It is therefore included in our dataset as a value of $\mb=2.07\pm1.03\times10^{10}$ \massunits.  We note that changes in the treatment of this measurement yield a much smaller difference in our combined results than our final estimated errors.
\\ \\
\textbf{\citet{Dwek95}:}  As mentioned above, this work constrains the mass of the Galactic bulge using photometry taken from the \cobe/\emph{DIRBE} observations, measuring $\mb=1.3\pm0.5\times10^{10}$ \massunits, where a Salpeter IMF has been assumed.  We convert this measurement to the Kroupa IMF by multiplying it by 1.62, derived using our definition of stellar mass and the assumptions made for the stellar models of \citet{BC03}.  Hence, we include this measurement in our dataset as a value of $\mb=2.11\pm0.81\times10^{10}$ \massunits.
\\ \\
\textbf{\citet{Freud98}:} This work favors a bar-to-disk ($B/D$) luminosity ratio of 0.33 based upon matching their Galactic model to infrared observations of the Milky Way from \cobe.  This model again includes a holed stellar disk.  We integrate this profile both with and without the hole implemented, under the assumption that the mass gained by the disk when excluding the hole was previously incorporated into the bar, in order to calculate a hole-less $B/D$ value.  We note that the results of this study are published as a function of $R_0$ in their Table 4, allowing us to interpolate from this table to obtain $B/D$ at a given $R_0$, instead of assuming a scaling relation.  Thus, for each MC simulation we include within our dataset a value of $\mb=\mdi\times(B/D)$, where $(B/D)$ represents the hole-less model bar-to-disk ratio corresponding to $R_{0,i}$.  We ascribe a 50\% error bar to this value, accounting for uncertainty in the mass-to-light ratios of the bar and disk.  Similarly to the \citet{Picaud04} estimate, we conservatively add in quadrature to this error estimate a value of 50\% of the holed-disk correction at $R_0=8.33$ kpc.  As a nominal value, we list $\mb=0.47\pm0.65\times10^{10}$ \massunits\ in Table \ref{table:mbulge}, the result from assuming our standard disk model values.  This overall error estimate is held constant for other values for the disk parameters.
\\ \\
\textbf{\citet{Dehnen98}:} This study provides a parameterized model of the mass distribution in the Galaxy fitted to several observational constraints.  They assume a single-exponential profile for the stellar disk, much like \citet{Bovy2013}.  Tables 3 \& 4 of their paper list the resulting bulge mass when varying a large number of parameters in the model.  Comparing models 2, 2a, and 2b we find that \mb\ in their model approximately scales as $\sim$$(R_0/8\mathrm{ kpc})^{1/2}$.  To provide the central value of this estimate we interpolate amongst \mb\ results from their models 1--4 using the value $L_d/R_0=0.26$ from our standard disk model detailed in \S\ref{sec:mstar_diskmodel}.  We estimate the random uncertainty in this value via propagation of errors: $\sigma(\mb)^2\simeq\frac{1}{4}\left[\mb(L_d+\sigma(L_d))-\mb(L_d-\sigma(L_d))\right]^2$.  In addition, we estimate the systematic error to be the standard deviation of the results from \citeauthor{Dehnen98}'s models 2c--i, which we add in quadrature to the random error.  Lastly, as this is a dynamical measurement, we remove the contribution of brown dwarfs to the bulge mass budget by multiplying by a factor of $0.94\pm0.02$.  Overall, this is included in our dataset a value of $\mb=0.65\pm0.38\times10^{10}$ \massunits.

\subsection{Setting a Prior on \textbf{\mb}} \label{sec:mstar_prior}
To place a prior on the mass of the Galactic bulge, we again consider the properties of spiral galaxies previously observed in the local universe.  Some spirals appear to have no bulge component, whereas in extreme cases the bulge-to-total ratio can be as large as $B/T\sim0.8$ \citep[e.g.,][]{Simien,Gadotti}.  Given our prior understanding of the total mass of the Milky Way to be of order $\sim$$5\times10^{10}$ \massunits\ \cite[e.g.][and references therein]{McMillan} we assume the mass of the Galactic bulge can be anywhere in the range $0\mathrm{-}4\times10^{10}$ \massunits\ with equal weighting.  This is represented by a flat distribution, such that 
\begin{equation} 
	P(\mb)=
		\begin{cases}
		\frac{1}{4}\times10^{-10},& \text{if } 0 \leq \mb \leq 4\times10^{10} \,\massunits \\
		0,		& \text{otherwise.}\\
		\end{cases}
\end{equation}
\begin{figure}
\centering
\includegraphics[trim=0.55in 1.25in 0.7in 0.55in, clip=true, width=1.0\columnwidth]{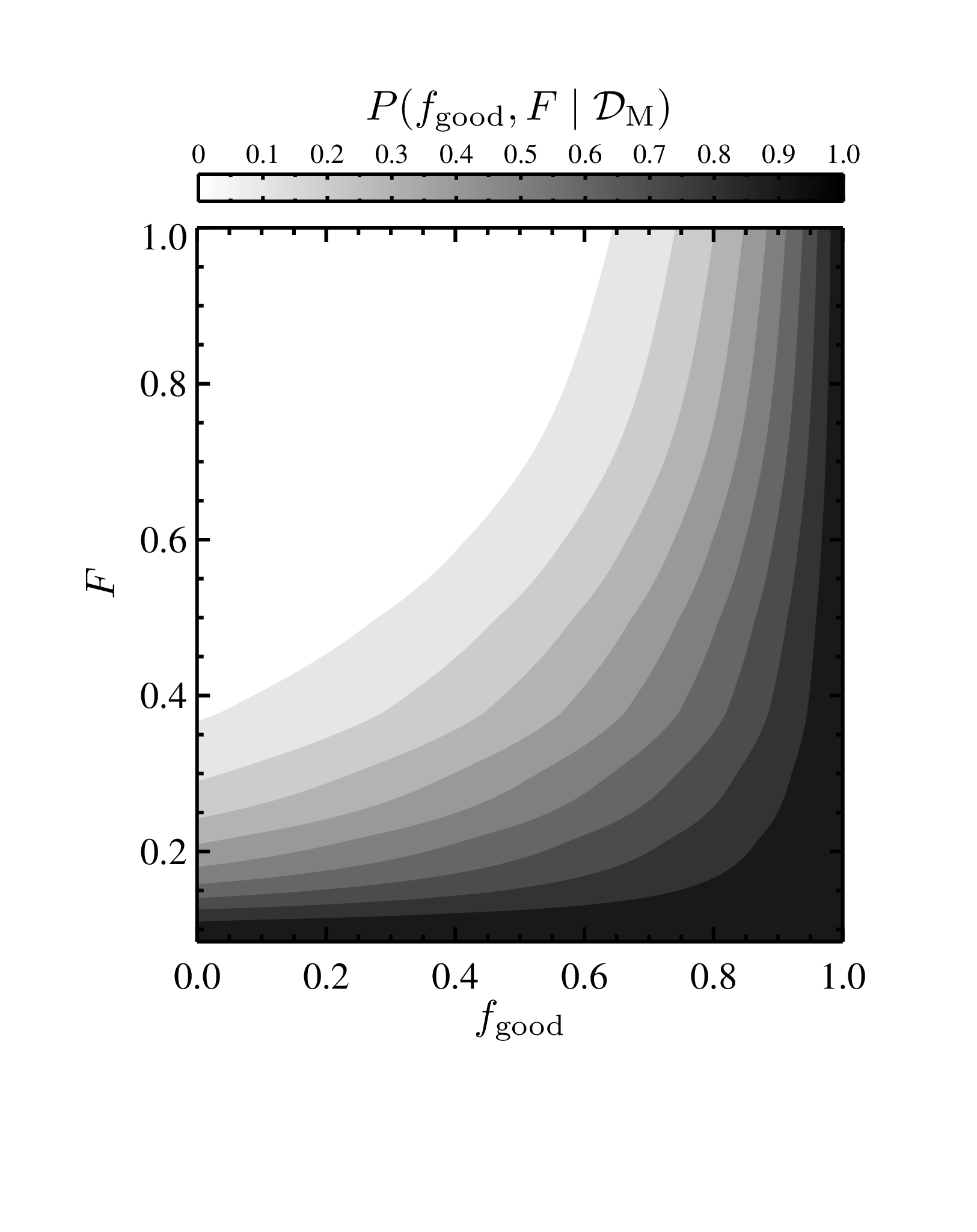}
\caption{Joint posterior PDF, $P({f_{\text{good}}, F \mid \D_\text{M}})$, for the free parameters describing the Milky Way bulge+bar mass dataset, $\D_\text{M}$, when applying a hierarchical Bayesian analysis.  The parameter $f_{\text{good}}$ quantifies the probability that any one measurement included in $\D_\text{M}$ is ``good'' (i.e., has accurately estimated error bars), and $F$ denotes the fraction of $10^{10}$ \massunits\ used as a floor on each individual error estimate when treated as not ``good'' in the model.  For ease of comparison, we have normalized the peak of this distribution to 1.  We see, similar to the results for the Milky Way SFR, that the likelihood, and thus the posterior, peaks near $f_{\text{good}}\approx1$ and $F\approx0$, indicating little tension amongst the Milky Way bulge+bar mass estimates.}
\label{fig:post_fg_F}
\end{figure}

\subsection{Stellar Mass Results} \label{sec:mstar_results}
\begin{figure}[t]
\centering
\includegraphics[trim=0.7in 1.25in 0.7in 1.5in, clip=true, width=1.0\columnwidth]{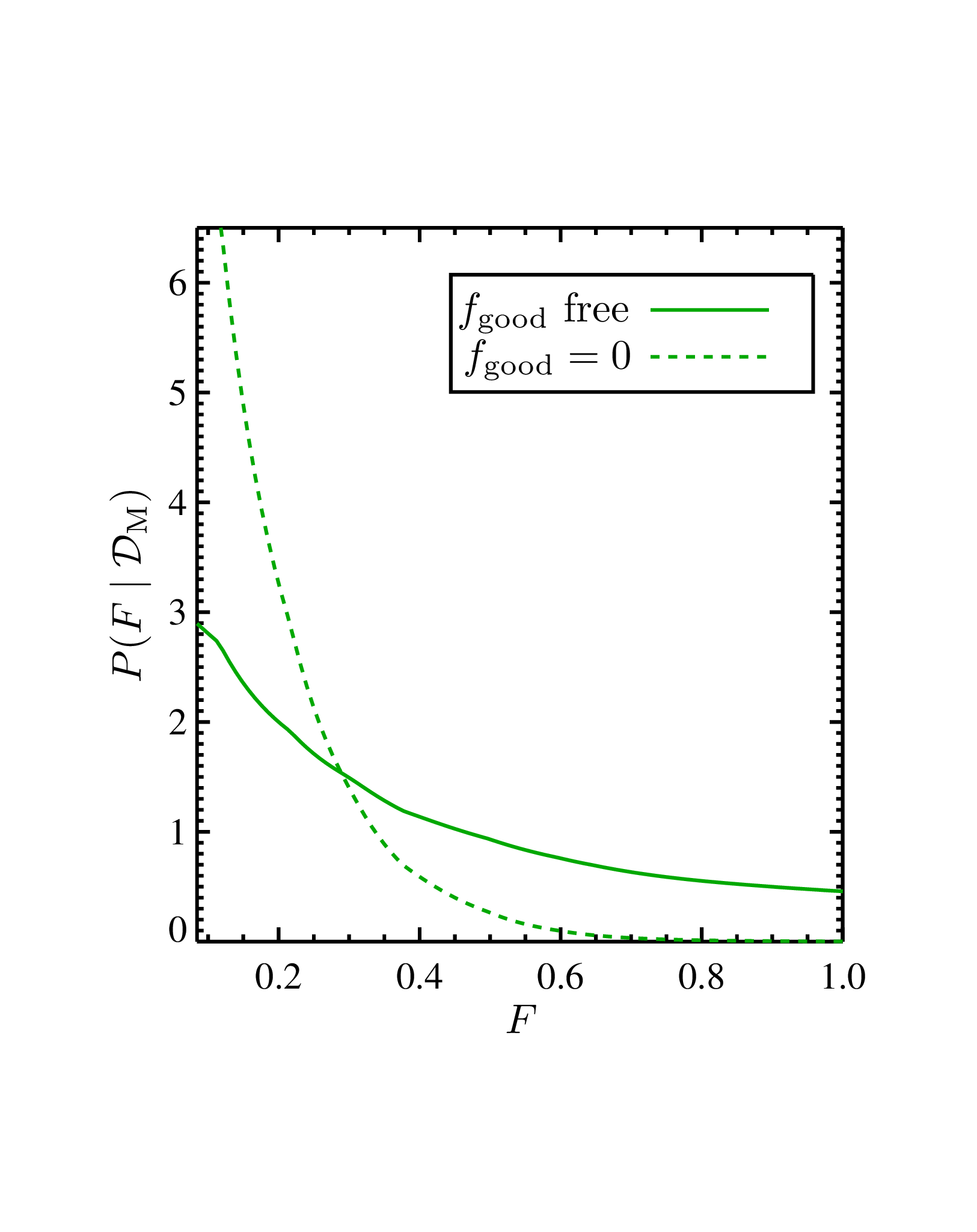}
\caption{Posterior PDF, $P({F \mid \D_\text{M}})$, for the floor value, $F$, defined as a fraction of $10^{10}$ \massunits\ imposed as a minimum error estimate for ``bad'' measurements in our Milky Way bulge+bar mass dataset, $\D_\text{M}$.  These results are obtained by marginalizing over all possible values of $f_{\text{good}}$ (solid curve) or else by setting $f_{\text{good}}=0$ (dashed curve).  Here we see the competition between multiple possible solutions within the HB analysis.  We note that both curves are marginalized over all possible values of the stellar mass contained in the bulge+bar component of the Galaxy, \mb.}
\label{fig:post_F}
\end{figure}

Table \ref{table:mbulge_results} shows the results for each model of ``bad'' measurements from our HB analysis.  We list the optimal value of $n$, $Q$, or $F$, as appropriate for each model, which we take to be the value corresponding to the peak of the marginalized posterior PDF for these parameters.  In Figure \ref{fig:post_fg_F}, we marginalize out \mb\ and show the resulting joint posterior for $f_{\text{good}}$ and $F$ normalized to a peak value of 1.
We can see in this case that the likelihood is maximized near $f_{\text{good}}\approx1$ and $F\approx0$, in similarity to the SFR results in \S\ref{sec:sfr_results} which favored $f_{\text{good}}\approx1$ and minimally adjusted error bars.  If we are to then marginalize out $f_{\text{good}}$ from the joint posterior, this yields the result $P({F \mid \D})$ shown in Figure \ref{fig:post_F} as a solid green curve.  Assuming that $f_{\text{good}}=0$ (i.e., that all included measurements are inaccurate to some extent) produces the the dashed green curve in Figure \ref{fig:post_F}; this is equivalent to the curve produced by cutting through the $f_{\text{good}}=0$ plane in Figure \ref{fig:post_fg_F}.
\begin{deluxetable*}{lcccccc}
\tablewidth{\textwidth}
\tablecaption{Combined \mb\ Results For Various Model Assumptions}
\tablehead{\multirow{2}{*}{Model} & Optimal & \mb$\pm1\sigma$ & \multirow{2}{*}{$k$\tablenotemark{b}} & \multirow{2}{*}{$\Delta$AIC\tablenotemark{c}} & \multirow{2}{*}{$\Delta$BIC} & \multirow{2}{*}{$\Delta\log_{10} K$} \\
& Value\tablenotemark{a} & (10$^{10}$ \massunits) & & & &}
\startdata
\\[-1.5ex]
\multicolumn{7}{c}{\underline{$f_{\text{good}}$ free --- Some of the measurements have inaccurate error bars.}} \\[1.5ex]
free-$n$ & 1.00 & $0.91\pm0.07$ & 3 & $4.00\pm0.00$ & $5.78\pm0.00$ & $-0.66\pm0.00$ \\
free-$Q$ & 0.00 & $0.91\pm0.08$ & 3 & $4.00\pm0.00$ & $5.78\pm0.00$ & $-0.45\pm0.00$ \\
free-$F$ & 0.08 & $0.92\pm0.08$ & 3 & $4.00\pm0.00$ & $5.78\pm0.00$ & $-0.42\pm0.00$ \\
$P_\text{bad}$ flat & N/A & $0.91\pm0.07$ & 2 & $2.00\pm0.00$ & $2.89\pm0.00$ & $-0.86\pm0.00$ \\[1.5ex]
\multicolumn{7}{c}{\underline{$f_{\text{good}}$=0 --- All of the measurements have inaccurate error bars.}} \\[1.5ex]
free-$n$ & 1.00 & $0.91\pm0.08$ & 2 & $2.00\pm0.00$ & $2.89\pm0.00$ & $-1.20\pm0.00$ \\
free-$Q$ & 0.00 & $0.93\pm0.09$ & 2 & $2.00\pm0.00$ & $2.89\pm0.00$ & $-0.83\pm0.00$ \\
free-$F$ & 0.08 & $0.94\pm0.10$ & 2 & $2.00\pm0.00$ & $2.89\pm0.00$ & $-0.85\pm0.00$ \\[1.5ex]
\multicolumn{7}{c}{\underline{$f_{\text{good}}$=1 --- None of the measurements have inaccurate error bars.}} \\[1.5ex]
all-``good''\tablenotemark{d} & N/A & $0.91\pm0.07$ & 1 & --- & --- & --- \\[1.5ex]
\multicolumn{7}{c}{\underline{Non--hierarchical combinations of the data.}} \\[1.5ex]
IVWM\tablenotemark{e} & & $0.88\pm0.06$ & \multicolumn{4}{l}{$\chi^2=14.03$} \\
IVWM($R_0=8.33$kpc) & & $0.91\pm0.06$ & \multicolumn{4}{l}{$\chi^2=13.69$}
\enddata
\tablecomments{The \mb\ results are well described by Gaussian distributions, and the values listed in this column represent the mean and 1$\sigma$ parameter of fits to these distributions.  The distributions we find for $\Delta$AIC, $\Delta$BIC, and $\Delta\log_{10} K$ from the MC simulations are strongly peaked but also highly asymmetric, and so in these columns we quote the median and standard deviation of the median obtained by bootstrapping these sets of values.  Errors of 0.00 indicate that the standard deviation of the median is $\ll$0.01.}
\tablenotetext{a}{The value of $n$, $Q$, or $F$ corresponding to the peak of marginalize posterior PDF for these quantities in each model.}
\tablenotetext{b}{The number of free parameters in the model.  See \S\ref{sec:better_model}.}
\tablenotetext{c}{For each iteration of the HB analysis, we record the AIC and BIC.  We then calculate the mean and standard error from the distribution of AIC and BIC values produced from all 10$^3$ iterations for each model.  $\Delta$AIC and $\Delta$BIC reflects the difference in the mean AIC and BIC value measured from the model which has the lowest AIC/BIC (here, this is the all-``good'' model).  We see 10$^3$ iterations yields sufficiently small standard errors (i.e., less than 0.5) in order to securely assess differences of 2, which would indicate a statistically significant difference between models.}
\tablenotetext{d}{Equivalent to the combined inverse-variance weighted mean (IVWM) from 10$^3$ MC simulations, uniformly scaling each estimate to the same $R_0$ value from our Galactic model (see Table \ref{table:params}).}
\tablenotetext{e}{The IVWM obtained from the nominal estimate values listed in Column 2 of Table \ref{table:mbulge}.  Essentially, since we have used flat priors, these are the same values we would find in the all-``good'' scenario if we did not use MC simulations to renormalize each estimate to the same choice of $R_0$.  IVWM($R_0=8.33$kpc) shows the results from the same calculation for the \mb\ estimates after scaling each to $R_0=8.33$ kpc, as in the last column of Table \ref{table:mbulge}.}
\label{table:mbulge_results}
\end{deluxetable*}

\begin{figure}
\centering
\includegraphics[trim=0.6in 1.25in 0.7in 1.5in, clip=true, width=1.0\columnwidth]{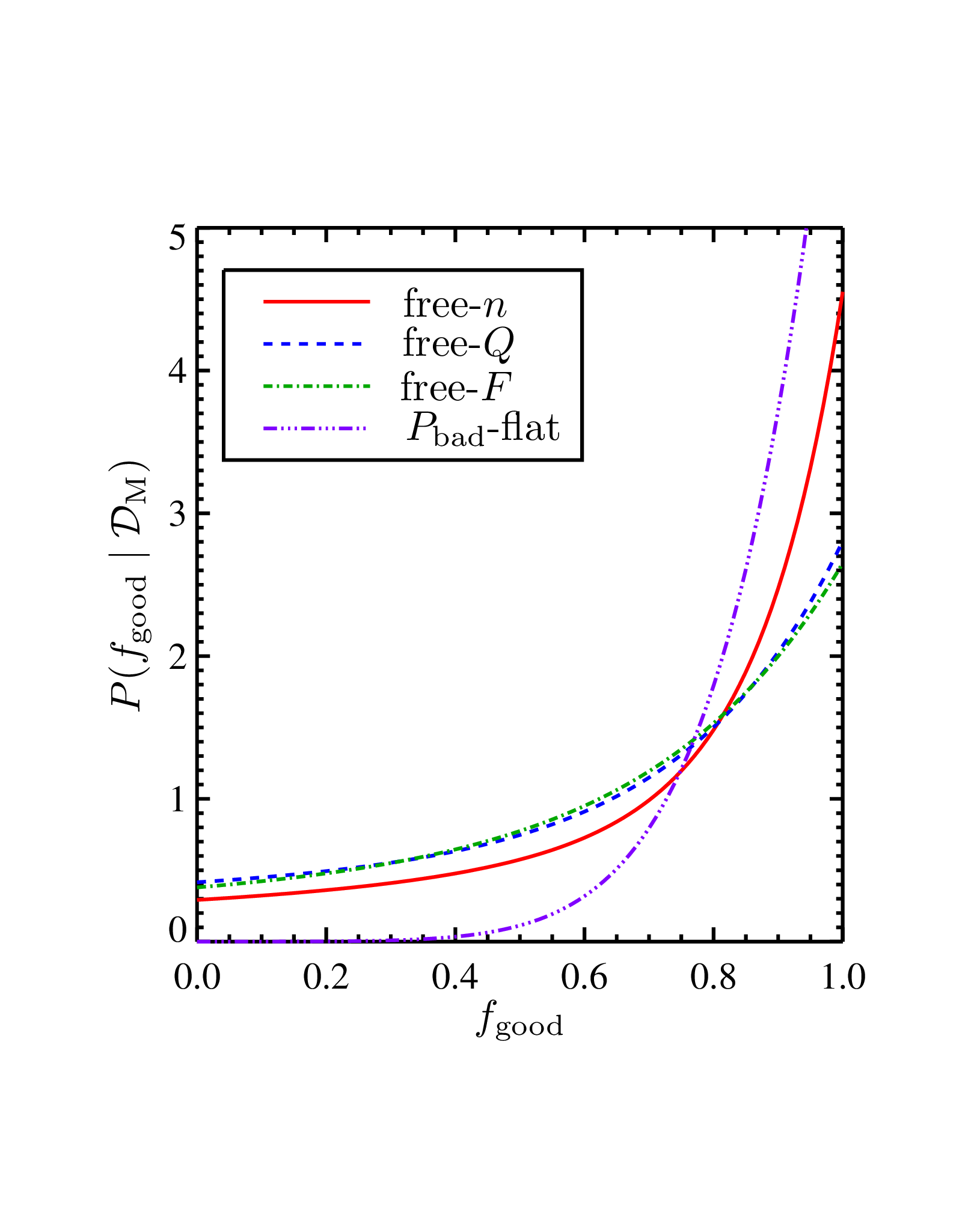}
\caption{Posterior PDF, $P({f_{\text{good}} \mid \D_\text{M}})$, for the fraction of ``good'' \mb\ measurements, $f_{\text{good}}$, (i.e., ones having accurately estimated errors) in our bulge+bar mass dataset, $\D_\text{M}$.  Each model shown accounts for the inclusion of ``bad'' measurements in a different way: the free-$n$ model (solid red) remedies underestimated errors by multiplying them by a scaling factor of $n$; the free-$Q$ model (dashed blue) adds extra error in quadrature to the errors on a measurement when treated as ``bad''; the free-$F$ model (dash-dotted green) places a floor on the errors of ``bad'' measurements; while the $P_\text{bad}$-flat model (triple-dot-dashed purple) works to completely disregard a measurement from $\D_\text{M}$ when considering it ``bad''.  Similar to the SFR results, the PDF for all ``bad''-measurement models are peaked around $f_{\text{good}}=1$, more strongly so for the free-$n$ and $P_\text{bad}$-flat scenarios than the free-$Q$ and free-$F$ scenarios.  Regardless, it turns out that the best model of $\D_\text{M}$, i.e., the one favored by the Bayesian evidence and information criteria, is one where $f_{\text{good}}$ is assumed to be 1, instead of allowing it as a free parameter -- i.e. an all-``good'' model.}
\label{fig:mb_fgood_comp}
\end{figure}

\begin{figure}
\centering
\includegraphics[trim=0.6in 1.25in 0.7in 1.5in, clip=true, width=1.0\columnwidth]{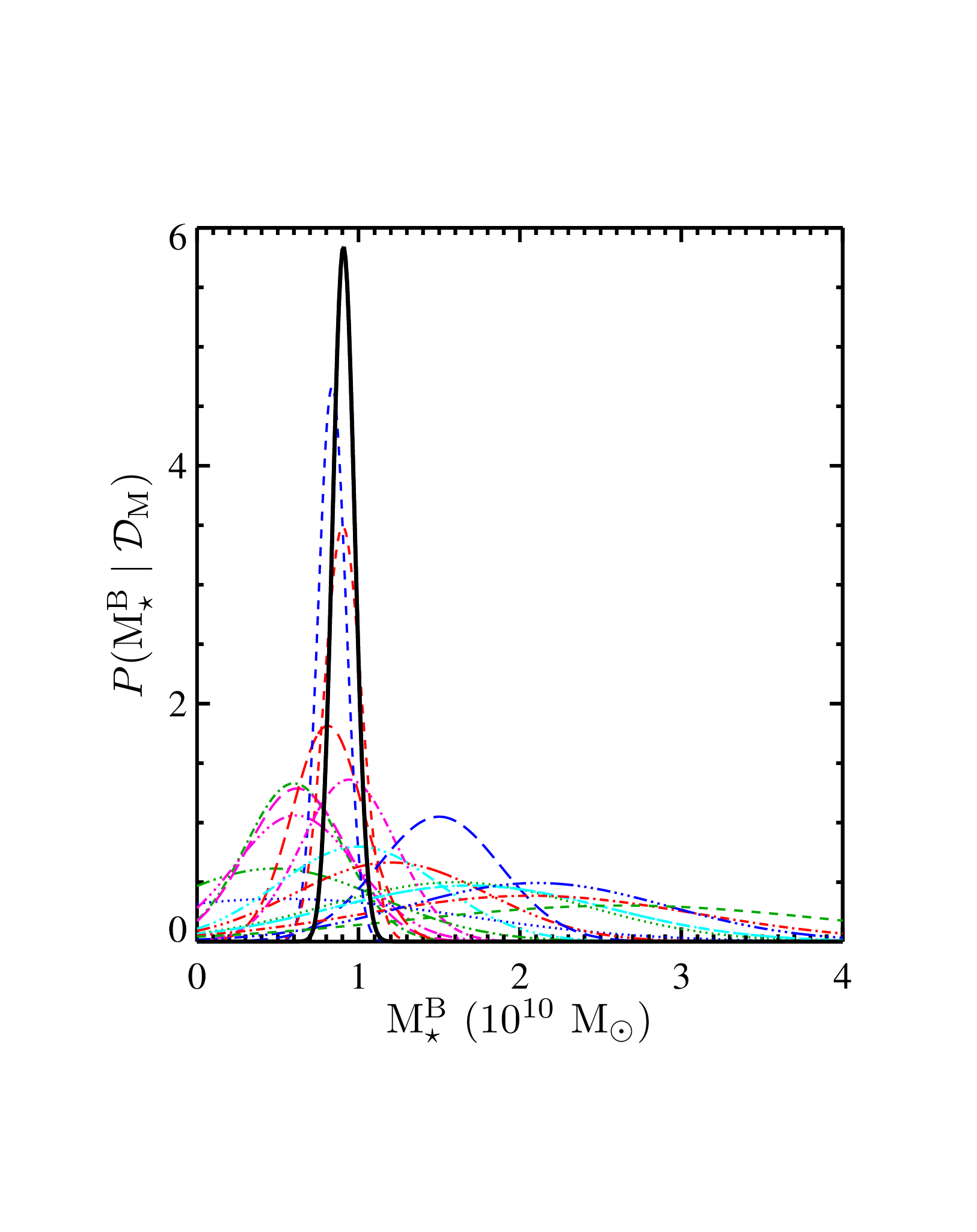}
\caption{Solid black curve shows the aggregate \mb\ result as determined from our HB analysis when using an all-``good'' model and Monte Carlo simulations to incorporate up-to-date information for the Galactocentric radius of the Sun, $R_0$, the scale length of the single-exponential disk, $L_d$.  Here, we have assumed $R_0=8.33\pm0.35$ kpc from the work of \citet{Gillessen} and $L_d=2.15\pm0.14$ kpc based on the measurements by \citet{Bovy2013}.  For comparison, we overlay the individual estimates from our Galactic bulge+bar mass dataset as dashed/dotted colored lines.  Some of these measurements are in tension with others; the HB methods allows us to account for that tension in obtaining consensus results.  Doubling the error bars on the \citet{Bissantz02} and \citet{Widrow08} \mb\ estimates (the relatively strongly-peaked dashed red and blue curves) yields only a 25\% increase in the error in our aggregate result, indicating that these estimates are not dominating in our final estimate.}
\label{fig:mb_comp}
\end{figure}

All model-comparison criteria listed in Table \ref{table:mbulge_results} result from comparing values with the all-``good'' model for each realization.  For example, $\Delta\log_{10} K$ for the free-$n$ model reflects the mean and standard error of the set $\left\{\log_{10}(K_{\text{free}-n}/K_{\text{all-``good''}})_i\right\}$, and similarly for $\Delta$AIC \& $\Delta$BIC.  In accordance with the likelihood peaking on the $f_{\text{good}}\approx1$ plane, as seen in Figure \ref{fig:post_fg_F} and \ref{fig:mb_fgood_comp}, all criteria values indicate that the best fit of the data results when the fewest free parameters are employed in the HB analysis.
That said, several of the other models listed in Table \ref{table:mbulge_results} have criteria values that do not differ enough to be statistically significant (i.e., their differences in $\Delta\log_{10} K$, $\Delta$AIC, or $\Delta$BIC are less than 2).  Again, similar to the SFR results, the model with highest $K$ is the one where all measurements are treated as accurate, indicating that once differences in the assumed Galactic model, IMF, and definition of \mt\ are accounted for, no correction for systematic effects is needed in order to relieve any tension in these measurements.  This is supported by the AIC \& BIC values as well.  Taking the all-``good'' model as our fiducial measurement, we therefore find the stellar mass of the Galactic bulge+bar to be $\mb=0.91\pm0.07\times10^{10}$ \massunits; a comparison with the other values in Table \ref{table:mbulge_results} shows that this estimate is quite robust to any way we account for ``bad'' estimates in the HB analysis.  We have also tested the impact that the \citet{Bissantz02} and \citet{Widrow08} \mb\ estimates make on our final results, as these two measurements are more strongly peaked than the other estimates and are centered at similar bulge mass values (see Figure \ref{fig:mb_comp}).  We find that doubling the nominal errors on both of these estimates produces only a 25\% increase in the error of our aggregate result and a negligible change to the mean value, indicating that these two estimates are not dominating our final estimate.

Lastly, we show in Table \ref{table:mbulge_results} the inverse-variance weighted mean (IVWM) of the nominal values from our \mb\ dataset (Table \ref{table:mbulge}) with no corrections to uniform $R_0$ and \md.  We can see that the IVWM is more strongly pulled by lower-valued outliers in comparison to the HB results.  This is likely in large part the result of the heterogeneous mixture of $R_0$ values in the sample, as the all-``good'' model is equivalent to 10$^3$ MC simulations of the IVWM, but rescaling all measurements to reflect $R_0=8.33\pm0.35$ kpc \citep{Gillessen}; in this case, the all-``good'' model corresponds to the average of the IVWM from our 10$^3$ MC simulations, rather than the IVWM of the set of original measurements.  With 18 \mb\ estimates in our dataset, and thus 17 degrees of freedom, we expect 68\% chance of finding $\chi^2$ in the range $[11.31,22.68]$ and a 95\% chance of it being in the range $[7.56,30.19]$.  Hence, the observed $\chi^2=14.03$ indicates only modest tension amongst the measurements; $\sim$2/3 of the time one would observe a $\chi^2$ this large or larger.

Similarly to our SFR analysis, we have tested whether common-mode systematics have a large impact on our bulge mass estimate, an effect our HB method is unable to account for as we assume that errors from each study are random compared to each other.  As initially discussed in \S\ref{sec:sfr_results}, this involves applying the HB method using the all-``good'' model to a bootstrap resampling of the \mb\ dataset in two different ways.  In the first case, we draw only one measurement at random from all those which use a common measurement technique (i.e., one of the four categorizations listed in Table \ref{table:mbulge} under the column labeled ``Constraint type''), yielding four estimates from a unique measurement method.  In the second case, we draw four measurements at random from the entire list of estimates in Table \ref{table:mbulge}.  For each case, we perform this process 1,000 times and measure the mean from the posterior distribution $P({\mb \mid \D})$ for each set of measurements.  We find the standard deviation of the mean to be 0.19 when resampling with only one measurement of each type from the \mb\ dataset and 0.23 when resampling at random from the entire dataset.  Again, there is greater scatter between measurements of the same type than between random measurements in the dataset.  Hence, similarly to our SFR results, we can safely conclude that common-mode systematics make a negligible impact on our bulge mass results.

The results from our HB analysis and MC simulations for $P({\md \mid \D})$ and $P({\mt \mid \D})$ are well-described by Gaussian distributions.  These correspond to $\md=5.17\pm1.11\times10^{10}$ \massunits\ and $\mt=6.08\pm1.14\times10^{10}$ \massunits.  We note that the mass of the stellar halo component of the Galaxy ($\sim$$10^9$ \massunits) is negligible compared to the uncertainties in our estimate \citep{Bell,Bullock}, and thus we can disregard its contribution.  The uncertainty in \mt\ is dominated by that in \md, making our \mt\ result highly insensitive to the choice of model used in the HB analysis of the Galactic bulge+bar mass estimates.  The largest sources of uncertainty in \md\ and \mt\ thus come from the values of $R_0$ and $L_d$ that we adopt (Table \ref{table:params}).  The constraints on these parameters are likely to improve with upcoming Galactic surveys, such as \textit{Gaia}, and so we also calculate derivatives of our mass estimates with respect to each so that they may be easily adjusted to reflect any improved information.  To do so, we simply redo our analysis after independently offsetting either parameter by one-half of its error estimate above and below the central value.  The derivative is calculated from the curve yielded from Lagrangian interpolation of the resulting 3 data points.  For convenience, we tabulate our stellar mass results for each component and their derivatives in Table \ref{table:derivs}.

Lastly, we display the PDF for the bulge-to-total mass ratio of our model in Figure \ref{fig:b2t} as a solid black curve, which indicates a median with 1$\sigma$ error estimate of $B/T=0.150^{+0.028}_{-0.019}$.  This is obtained from the distribution of values, $(B/T)_i=\mbi/(\mbi+\mdi)$, resulting from each self-consistent realization of the Galaxy.  Calculating $P(B/T)$ in a model-consistent way (i.e. accounting for covariances within our model) yields measurably tighter constraints than when doing so by combining independent estimates of \mb\ and \mt.  For example, the blue dashed curve in Figure \ref{fig:b2t} is the result of combining randomly drawn pairs of \mb\ and \md\ values drawn from the PDFs for each mass component in this study and assuming $\mt=\mb+\md$.  The red dash-dotted curve in Figure \ref{fig:b2t} shows the distribution of bulge-to-total luminosity ratios for a sample of 212 Sbc-Sc galaxies measured from SDSS by \citet{Oohama}.  The mass-to-light ratio of the older stellar population in the bulge is higher than that of the younger stellar disk and so converting this to a bulge-to-total mass ratio would shift the distribution toward slightly higher values; however, we can safely assess that Milky Way lies well within the range of $B/T$ values.  \citet{GrahamWorley} find the average bulge-to-disk $K$-band flux ratio for a sample of 79 Sbc galaxies to be $\log(B/D)=-0.82$, which converts to an average $B/T$ of 0.13; this compares well with our Milky Way value.
\begin{deluxetable*}{cccc}
\tablewidth{\textwidth}
\tablecaption{Milky Way Properties and Derivatives}
\tablehead{Property & Fiducial Result & $\partial/\partial R_0$ & $\partial/\partial L_d$}
\startdata
\mb & $0.91\pm0.07\times10^{10}$ \massunits & $0.093\times10^{10}$ \massunits\ kpc$^{-1}$ & $0.004\times10^{10}$ \massunits\ kpc$^{-1}$ \\
\md & $5.17\pm1.11\times10^{10}$ \massunits & $3.000\times10^{10}$ \massunits\ kpc$^{-1}$ & $0.469\times10^{10}$ \massunits\ kpc$^{-1}$ \\
\mt & $6.08\pm1.14\times10^{10}$ \massunits & $3.093\times10^{10}$ \massunits\ kpc$^{-1}$ & $0.473\times10^{10}$ \massunits\ kpc$^{-1}$ \\
$B/T$ & $0.150^{+0.028}_{-0.019}$ & $-0.061$ kpc$^{-1}$ & $-0.011$ kpc$^{-1}$ \\
SFR & $1.65\pm0.19$ \sfrunits & 0 \sfrunits\ kpc$^{-1}$ & 0 \sfrunits\ kpc$^{-1}$ \\
sSFR & $2.71\pm0.59\times10^{-11}$ yr$^{-1}$ & $-1.381\times10^{-11}$ yr$^{-1}$ kpc$^{-1}$ & $-2.111\times10^{-12}$ yr$^{-1}$ kpc$^{-1}$
\enddata
 \tablecomments{Derivatives of \mt\ are not simply the sum of the derivatives for \mb\ and \md\ due to covariances between the parameters in our model.  Derivatives of $B/T$ represent the change in the median bulge-to-total mass ratio with respect to the appropriate parameter, measured independently from any other quantity in this table.  The sSFR and its derivatives, however, are calculated directly from the \sfr\ and \mt\ results; e.g., here $\partial(\sfr/\mt)/\partial R_0=-\sfr\,\text{M}_\star^{-2}\times\partial\mt/\partial R_0$ since $\partial\sfr/\partial R_0=0$.}
\label{table:derivs}
\end{deluxetable*}

\section{Summary \& Discussion} \label{sec:summary}
In this paper we have developed improved constraints on several of the Milky Way's global properties.  We build upon the prior measurements found in the literature, joining them into consensus results using the power of the hierarchical Bayesian (HB) method.  This method \citep{Press,LangHogg} takes into account the possibility of inaccurate measurements being included in our datasets, and has proven to be quite robust when varying how we dealing with such ``bad'' measurements.  By incorporating all the information contained in the individual measurements, we have obtained significantly improved constraints on the Milky Way properties we investigate.  At the same time, given the expectation that there could be systematics affecting the Milky Way data, the HB method has given us confidence in the robustness of our results, even though Occam's razor favored the simplest model (i.e., the inverse-variance weighted mean) in the end.  For convenience, we tabulate the main results from this study in Table \ref{table:derivs}, as well as their derivatives with respect to the Galactocentric radius of the Sun, $R_0$, and the exponential scale length of the disk, $L_d$, allowing for our results to be updated if constraints on these parameters improve.

In our first application, we capitalize on the work of \citet{Chomiuk}, who provide a tabulation of the star formation rate (SFR) estimates in the literature over the last several decades.  The authors make huge strides in placing each measurement on equal footing by renormalizing them all to the same choice of initial mass function and stellar population synthesis model.  Using these updated estimates as our dataset, we find the SFR of the Milky Way to be $\sfr=1.65\pm0.19$ \sfrunits\ (assuming a Kroupa IMF and Kroupa-normalized Kennicutt ionizing photon rate).

Next we investigate the stellar mass contained in each of the major components of the Milky Way.  In Table \ref{table:mbulge}, we have compiled an extensive list of Galactic bulge, pseudo-bulge, and/or bar mass estimates from the literature.  We assume the single-exponential density profile for the disk laid forth by \citet{Bovy2013}; we tabulate the adopted probability distributions for all relevant parameters in Table \ref{table:params}.  We then combine our HB analysis with Monte Carlo (MC) simulations that uniformly scale each estimate to the same value of $R_0$, ensuring propagation of $R_0$ errors into the hierarchical \mb\ result.  The MC calculations also yield estimates of the stellar mass of the bulge+bar, the stellar mass of the disk, and the total stellar mass in the Milky Way.  

Our combined estimate of the bulge+bar mass is $\mb=0.91\pm0.07\times10^{10}$ \massunits.  This estimate has substantially smaller errors than the individual estimates used to derive it.  We note that the error given not only reflects the random and systematic uncertainties in each individual estimate, but also current uncertainties in $R_0$.  Our results show that once we have renormalized each estimate to reflect the same choice of IMF, stellar distribution profile for the disk, and definition of stellar mass that likely there are minimal systematics to further correct for.  We adopt the Kroupa IMF in this study and include the contributions of main-sequence stars and remnants, but not substellar material, in our definition of stellar mass in accord with the MPA-JHU measurements of SFR and \mt\ for external galaxies found in SDSS.

Under consistent assumptions, we find the disk mass of the Milky Way is $\md=5.17\pm1.11\times10^{10}$ \massunits.  Our model of the Galactic stellar disk is based on that of \citet{Bovy2013}, but we directly incorporate current uncertainties in the Galactocentric radius of the Sun.  Our disk mass estimate is in broad agreement with other measurements found in the literature.  One of the earliest models of the Milky Way by \citet{Bahcall}, primarily constrained by star counts from photometric observations of the Galaxy, yields a stellar disk mass of $5.6\times10^{10}$ \massunits.  The $\Lambda$CDM-based Milky Way models by \citet{KZS02} favor a range of $4\mathrm{-}5\times10^{10}$ \massunits.  Similarly, the kinematically constrained models from \citet{Dehnen98} find the mass of the disk to lie in the range of $4.2\mathrm{-}5.1\times10^{10}$ \massunits.

We find that total stellar mass in the Milky Way is $\mt=6.08\pm1.14\times10^{10}$ \massunits\ by statistically combining the stellar disk and bulge+bar mass estimates through model-consistent MC simulations (the mass of the stellar halo component ($\sim$10$^9$ \massunits) is negligible compared to our uncertainty).  The overall error in our \mt\ estimate is dominated by that contributed from the disk component, rendering our final result highly insensitive to any assumptions involved with combining bulge mass estimates to determine $P({\mb \mid \D})$.  In comparison, \citet{McMillan} find a total baryonic mass of $6.43\pm0.63\times10^{10}$ \massunits, from which we can subtract the atomic and molecular phase gas mass of $9.5\pm3.0\times10^9$ \massunits\ (\citet{Dame}, corrected for helium contributions by \citet{Flynn}), giving $\mt=5.48\pm0.70\times10^{10}$ \massunits.  In addition, \citet{Flynn} give a back-of-the-envelope calculation for \mt, finding it to be in the range $4.85\mathrm{-}5.5\times10^{10}$ \massunits.  Our estimate for the total stellar mass compares well with the results of these two recent studies, but takes advantage of a large sample of bulge mass measurements found in the literature and is founded on improved knowledge of several of the properties of the Galactic disk from SDSS.  Again, we stress that all of our mass results assume a Kroupa IMF and an exponential profile for the Galactic disk, to match the assumptions used in studies of extragalactic objects.
\begin{figure}
\centering
\includegraphics[trim=0.6in 1.25in 0.7in 1.5in, clip=true, width=1.0\columnwidth]{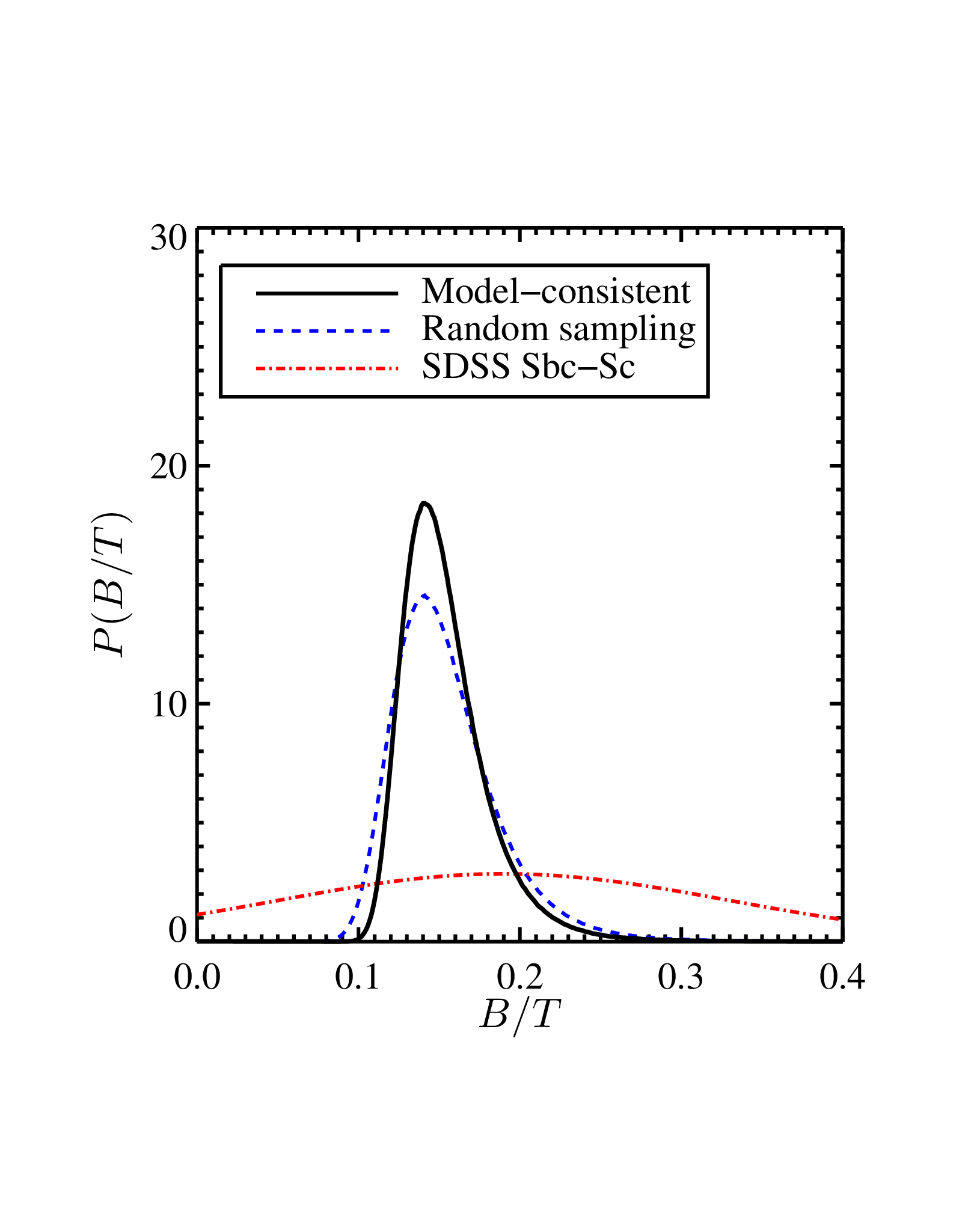}
\caption{Solid black curve shows the PDF for the bulge-to-total mass ratio, $B/T$, of the Milky Way as determined by our HB analysis.  This is produced through Monte Carlo simulations where each realization of the bulge+bar mass (\mbi) and disk mass (\mdi) are determined using the same values for structural parameters for the disk, $\left\{R_0,L_d,\Sigma_\star(R_0)\right\}_i$, randomly drawn from their respective distributions listed in Table \ref{table:params}.  The histogram of values $(B/T)_i=\mbi/(\mbi+\mdi)$ is then normalized to integrate to unity.  The dashed blue curve shows the result of simply drawing from the fiducial estimates of $P({\mb \mid \D})$ and $P({\md \mid \D})$ from our model independently.  Calculating $B/T$ in a model-consistent manner yields a noticeably tighter constraint than when not accounting for covariances between \mb\ and \md.  Lastly, we overlay the distribution of $B/T$ luminosity ratios for a sample of 212 Sbc-Sc galaxies measured from SDSS by \citet{Oohama}.  Although converting to $B/T$ mass ratios would produce a slight shift toward higher values, we see the $B/T$ ratio for the Milky Way is not unusual for galaxies of its Hubble type.}
\label{fig:b2t}
\end{figure}

Ultimately, the constraints we are able to place on the stellar mass of the disk component as well as the total stellar mass of the Galaxy depend strongly on the value we assume for $R_0$.  For our main results, we have chosen the \citet{Gillessen} estimate of $8.33\pm0.35$ kpc for its direct determination of the distance to Sgr A* (i.e., the center of the Milky Way) based on the orbits of nearby stars, its thorough treatment of systematic errors, and (thanks in part to the breadth of its errors) its consistency with a large variety of other $R_0$ measurements (both direct and indirect) in the literature, which range as low as $\lesssim$8 kpc or as high as $\gtrsim$8.5 kpc.  It is likely, however, that constraints on $R_0$ will tighten as geometric methods are continually improving, while indirect methods that are plagued by systematic errors become obsolete \citep{Genzel}.  For instance, the most recent measurement by \citet{Chatzopoulos} finds $R_0=8.30\pm0.09|_\text{stat}\pm0.10|_\text{syst}$ kpc by dynamically modeling the nuclear star cluster dynamics. The constraints from that analysis in the distance vs. mass of the black hole plane are in modest tension with those from monitoring stellar orbits by \citet{Gillessen}; a joint analysis of the two ignoring this tension yields $R_0=8.36\pm0.11$ kpc.  If we are to use this result for our study, the uncertainty in both the disk mass and the total mass are reduced to $0.5\times10^{10}$ \massunits, and our bulge mass error is reduced to $0.06\times10^{10}$ \massunits, while the central values are changed only slightly.

From the distribution of model-consistent realizations of \mb\ and \md, accounting for covariance between the two, we find that the Milky Way has a bulge-to-total mass ratio of $B/T=0.150^{+0.028}_{-0.019}$.  As seen in Figure \ref{fig:b2t}, this result makes our Galaxy typical amongst galaxies of similar morphological type in the local universe.  Finally, combining our results for \sfr\ and \mt, we find that the specific star formation rate of the Milky Way is $\sfr/\mt=2.71\pm0.59\times10^{-11}$ yr$^{-1}$.

In a paper soon to follow, we will continue towards our goal of producing a better global picture of the Milky Way.  With the improved constraints placed on \mt\ and \sfr\ in this work, we will next show that we can convert this information into accurate predictions for the integrated photometric properties of the Milky Way; i.e., the brightness and color of our Galaxy as they would be observed by alien astronomers from across cosmological distances.  All of this work will culminate in a newfound ability to accurately place the Milky Way in context -- i.e., we now have tight constraints on where our Galaxy falls compared to observational trends we find for other galaxies.

\section*{Acknowledgements}
We are grateful to Simon Mutch, Harry Ferguson, Sandra Faber, Leo Blitz, Jo Bovy, Hongsheng Zhao, Renbin Yan, Rachel Somerville, and Alister Graham for helpful discussions.  We thank Chad Schafer especially for a detailed reading and comments, as well as assistance with clarifying the strict-prior assumptions used in section \ref{sec:monte_carlo}.  We also thank the anonymous referee for leading us to a number of improvements in this work.  TCL and JAN are supported by the National Science Foundation (NSF) through grant NSF AST 08-06732.

\footnotesize
\bibliographystyle{apj}
\bibliography{mstar_sfr_paper}

\normalsize
\appendix
\section{Alternative models of the Galactic disk} \label{sec:alt_disk_models}
If the exponential disk model used by \citet{Bovy2013} is far from reality, the total mass estimates in this paper will, of course, be incorrect.  The true global structure of the Galactic disk remains an active area of research, and the direct measurements available in the literature that probe the (luminosity and mass) distribution of its stars have generally been limited to the range of $3\lesssim R\lesssim9$ kpc \citep[e.g.,][]{Juric,Bovy2013}.  Limiting the range of radii in this way mitigates the problem of needing to distinguish bulge/bar stars from disk stars where they overlap.  These components may be separated with kinematical information, but that is available for few stars and is of course impossible for studies based on aggregate light \citep[e.g.,][]{Freud98,DrimmelSpergel01}.  Similarly, studies that investigate the inner stellar mass at $R\lesssim3$ kpc, such as all of those in our Table \ref{table:mbulge}, must account for the bulge+bar and disk components either by fitting for them simultaneously or subtracting off the contribution from stars in the inner disk based on the model that is assumed.  We remind the reader that, as described in \S\ref{sec:mstar_data}, we have attempted to renormalize all of the bulge+bar mass measurements in Table \ref{table:mbulge} to reflect uniform assumptions about the disk.  Ultimately, all of this data is consistent with an exponential mass density profile, such as the one we have used for this study; in general, structural decompositions and model magnitude measurements for external galaxies also assume such a profile for disks.

In order to assess the impact which a non-exponential mass profile would have on the inferred total disk mass, we investigate the impact of allowing the mass profile to be described by a S\'ersic model \citep{Sersic68} in the radial direction, rather than a pure exponential, which in the radial (but not azimuthal) direction is equivalent to a S\'ersic profile with an index of $n=1$.  S\'ersic indices from global fits to the light from star-forming galaxies in the Sloan Digital Sky Survey cluster around a value just above 1, with a tail to larger values whose strength increases to redder colors \citep{Blanton2003}.  This is consistent with a picture where disks are indeed exponential and larger S\'ersic indices are obtained in earlier-type spirals with greater bulge contributions (a de Vaucouleurs-profile bulge would have a Sersic index of $n=4$).

To investigate this further, we have employed the NYU Valued-Added Galaxy Catalog \citep[VAGC;][]{Blantonsersic}, which includes S\'ersic radial profile fits to the $r$-band 2D images of a sample of galaxies from the Seventh Data Release \citep[DR7;][]{DR7} from the Sloan Digital Sky Survey \citep{York2000}.  These fits provide the S\'ersic index ($n$), scale radius ($r_0$), and the covariance matrix between $n$ and $r_0$.  From Data Release 8 \citep[DR8;][]{DR8} we have also obtained the spectroscopically measured redshift ($z$), the fraction of light from a de Vaucouleurs (bulge-like) profile when combined with a pure exponential ($n=1$) profile that best fits the 2D image (\texttt{fracDeV}), the axis ratio ($b/a$) from the best-fit exponential profile to the 2D image, and the $g-r$ color from \texttt{model} magnitudes that are extinction- and $K$-corrected to $z=0$ \citep[which we denote $^0(g-r)$;][]{kcorrect}.  We restrict to only those galaxies found in both the NYU-VAGC and DR7/8 datasets that are also part of the cleanly-measured volume-limited sample described in \citet{Licquia15}.  Overall, this ensures that all galaxies lie in the range $0.03<z<0.09$ and have cleanly measured images.  From this sample we then reduce to only those that meet the following criteria: $^0(g-r)<0.5$, $\sigma(n)<0.25$, $\sigma(r_0)<0.25$, $\texttt{fracDeV}<0.1$, and $b/a>0.7$.  This yields a set of 5,533 star-forming galaxies with minimal bulge-like components that appear $\sim$face-on and that have well-measured radial S\'ersic profile fits.  We expect that the S\'ersic indices for this sample can be used to broadly constrain the plausible values for the Galactic disk.  

We find the median of this distribution to be 1.08 and the 1$\sigma$ range to be $[0.9,1.3]$.  Given that this sample still includes objects with a nonzero bulge mass (even a small de Vaucouleurs bulge would increase the combined S\'ersic index), values of $n$ near but larger than 1 can be easily explained by bulge contributions to the light.  Apparent values of $n$ below 1 are predicted to be observed for intrinsically pure exponential disks due to projection effects and dust absorption \citep[see][and references therein]{Pastrav13a,Pastrav13b}.  Furthermore, one expects substantial scatter when fitting S\'ersic models to objects with visible substructure (e.g., spiral arms).  As a result, the S\'ersic index distributions of disk-dominated galaxies in SDSS appear to be highly consistent with a scenario where the effective S\'ersic index for disks, if measured with no bulge contribution, would in fact be $n=1$ (i.e., exponential).  We therefore do not find any compelling evidence that would cause us to discard the assumption that the Milky Way disk is exponential so that a more complicated model is needed.

Although we have not found any compelling evidence that the exponential disk assumption should be discarded, in order to assess what the impact of non-exponential profiles could be on our results, we have explored the effect that varying the S\'ersic index has on the distribution of stellar mass in the Milky Way compared to a pure exponential ($n=1$) profile, while keeping fixed the mean values of the disk parameters listed in Table \ref{table:params}.  For fixed disk parameters, decreasing values of $n$ increasingly add more stellar mass towards the center of the Galaxy, while increasing values of $n$ remove it.  For instance, decreasing from $n=1$ to 0.9 effectively $\sim$doubles $\Sigma_\star$ at the the Galactic center as well as the total disk mass at $R<3$ kpc (from $1.97\times10^{10}$ \massunits\ to $3.74\times10^{10}$ \massunits), which would imply that the disk is the dominant component over the bulge at all radii.  The possibility that the disk may be described by a radial S\'ersic index that is non-negligibly below 1 can therefore be discarded, as such models would be in strong tension with direct measurements of the Galactic rotation curve at these inner radii, as well as the mass measurements of the bulge we use in this study.

In contrast, we find that the mass of the disk between $R=3$ kpc (roughly where the bulge truncates) and $R=25$ kpc (well beyond the likely truncation radius of a S\'ersic-model disk) is rather stable for $n\geq1$; increasing $n$ anywhere from 1 up to 1.5 yields changes that are always $\lesssim5\times10^9$ \massunits, which is modest compared to the overall uncertainties in the disk mass estimate we present in \S\ref{sec:mstar_results}.  For changes at $R<3$ kpc, disk mass and bulge/bar mass are traded off against each other (due to our definition of the central mass), so total masses are minimally affected.  Of course, given that we have renormalized all of the bulge+bar mass measurements in Table \ref{table:mbulge} to reflect an exponential ($n=1$) disk, it would not be consistent to combine our HB bulge+bar mass result with a generalized S\'ersic model of the disk.  Hence, the effects we have explored here only constitute a first-order correction; doing better would require determining how a wide variety of historical bulge mass measurements would change if a very different disk model were used in fitting, which goes far beyond the scope or purpose of this paper.  We note that, since our definition of the bulge/bar mass is the additional mass in excess of an extrapolated exponential disk, even if the disk equivalent $n$ were in fact less than 1, we would have included that mass in our bulge component, so our total mass would be minimally affected (with only the fraction of that mass assigned to the bulge and disk changing).

We note that the ultimate goal of this study is to produce measurements of the mass of the Milky Way that may be directly compared to those for other galaxies, particularly the mass determinations from the MPA-JHU SDSS catalog.  The photometry in that catalog is based on model fits to galaxies that assume de Vaucouleurs bulges and pure exponential disks; if a very different model were used for determining the Milky Way mass, it is unlikely the results would be directly comparable.  We also note that studies of the Galactic disk find that its vertical structure is well described by an exponential distribution with roughly constant scale height \citep[e.g.,][]{Juric,Bovy2013}, inconsistent with a S\'ersic model.  Given all this, we consider an exponential mass profile, rather than a generalized S\'ersic profile, the best option for modeling the Galactic disk.  Should this assumption prove incorrect in the future (e.g., based on additional tests that will be provided by \textit{Gaia} \citep{Gaia} or APOGEE-2 \citep{apogee} measurements), nonetheless we have found that changing the disk model within reasonable bounds should have negligible impact on our total stellar mass estimates.

\end{document}